# Current state and perspectives of nanoscale molecular rectifiers

*Ritu Gupta[1], Jerry A. Fereiro[2]\*, Akhtar Bayat[3], Michael Zharnikov[4]\*, and Prakash Chandra Mondal[1]\**

[1]Department of Chemistry, Indian Institute of Technology Kanpur, Kanpur-208016, India

[2]School of Chemistry, Indian Institute of Science Education and Research Thiruvananthapuram, Thiruvananthapuram-695551 Kerala, India

[3]Laboratoire Photonique Numérique et Nanosciences, UMR 5298, Université de Bordeaux, 33400 Talence, France

[4]Angewandte Physikalische Chemie, Universität Heidelberg, Im Neuenheimer Feld 253, 69120 Heidelberg, Germany

E-mail: jerryfereiro@iisertvm.ac.in (J.A.F); Michael.Zharnikov@urz.uni-heidelberg.de (M.Z.); pcmondal@iitk.ac.in (PCM)

**Abstract:** The concept of utilizing a molecule bridged between two electrodes as a stable rectifying device with the possibility of commercialization is a "holy grail" of molecular electronics. Molecular rectifiers do not only exploit the electronic function of the molecules but also offer the possibility of their direct integration into specific nano-electronic circuits. However, even after nearly three decades of extensive experimental and theoretical work, the concept of molecular rectifiers still has many unresolved aspects concerning both the fundamental understanding of the underlying phenomena and the practical realization. At the same time, recent advancements in molecular systems with rectification ratios exceeding $10^5$ are highly promising and competitive to the existing silicon-based devices. Here, we provide an overview and critical analysis of the current state and recent progress in molecular rectification relying on the different design concepts and material platforms such as single molecules, self-assembled monolayers, molecular multilayers, heterostructures, and metal-organic frameworks and coordination polymers. The involvement of crucial parameters such as the energy of molecular orbitals, electrode-molecule coupling, and asymmetric shifting of the energy levels will be discussed. Finally, we conclude by critically addressing the challenges and prospects for progress in the field and perspectives for the commercialization of molecular rectifiers.

## 1. Introduction

As the size of the electronic components in modern-day devices approaches the few-nanometer (nm) limit, complementary metal-oxide-semiconductor (CMOS) elements start to face enormous complexities owing to issues such as short channel effects, the thickness of oxide layers, and lithography limitations. In this context, considerable efforts have been expended over the last three decades to explore the feasibility of using molecules, with the sizes on the order of few nm, as an alternative to CMOS circuit elements. This idea is backed by the fact that essentially all electronic processes in nature, from photosynthesis to signal transmission, are mediated by molecules and their assemblies. An extension of conventional electronics to molecular building blocks, connected to the real world by suitable electrodes, would not only allow further miniaturization of electronic devices but also provide a valuable opportunity of exploring intrinsic properties of molecules at the nm scale[1–3]. Prototype systems in this context are two-terminal molecular junctions that are now frequently used as a testbed for investigating structure-function relationship in passive and stimuli-responsive charge transport[4–6]. Such a relationship represents a key advantage of molecular electronics, on the ground of the assumption that a function of a molecular device can be controlled by chemical synthesis to degrees that are difficult to achieve within the standard, solid-state CMOS technology. The respective rational design should ensure that specific, predefined electronic functions of such a device are primarily determined by the nature and electronic characteristics of the molecules, making them truly molecular devices[7–10]. Currently, this relationship is sometimes not entirely straightforward, which eventually casts into doubt that some of the reported "molecular devices" are, in reality, truly molecular. Further requirements for a working molecular electronic device include its stability, reliability, reproducible performance, high yield, and the possibility of its integration with other elements of an electronic circuit.

Meanwhile, the concept of molecular electronics, which was initially something between science fiction and wishful thinking, has been experimentally realized[11–14], with the availability of different design frameworks and a variety of sophisticated fabrication and analysis techniques[15–20]. The whole family of basic electronic components, such as resistors, capacitors, diodes, transistors (still a challenge), and memory elements, can be realized by embedding molecules between two conducting electrodes and contacting them, if necessary, by an additional electrode or an external stimuli[21,22]. Among these components, molecular diodes are one of the most straightforward, owing to their relative simplicity and usefulness, in combination with other building blocks, for various functions in electronic circuits. A molecular diode acts electronically, very similar to its semiconductor counterpart, allowing the current to flow in one direction and restricting the flow in the opposite direction, i.e., a molecular diode behaves as a *rectifier*. The performance of a rectifier, in general, is quantified by the rectification ratio (*RR*), which represents the ratio of the current density (*J*) at a given forward bias ($+V_x$) compared to the current density at the reverse bias of the same magnitude ($-V_x$). Accordingly, $RR(V_x) = |J(+V_x)/J(-V_x)|$ with the sign of the bias being usually chosen to yield *RR* >1. Following the initial theoretical proposition of one-molecule rectifier by Aviram and Ratner,[23] many claims of molecular rectification using different molecules, diverse fabrication strategies, and various analysis platforms have been reported. Few representative examples of different molecular designs applied currently in context of rectification are compiled in Table 1.

**Table 1. Different types of molecular design in context of rectification**

| Type of Molecules | Structure of molecules | References |
|---|---|---|
| **D-σ-A Type** | 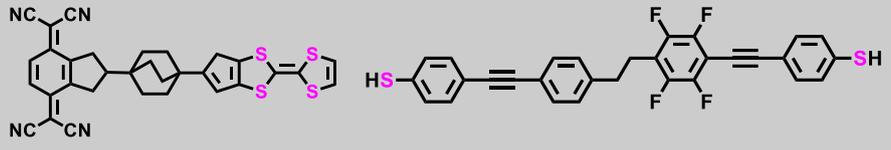 | 23,24 |
| **D-A Type** | 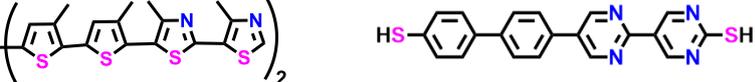 | 25,26 |
| **D-Perpendicular-A Type** | 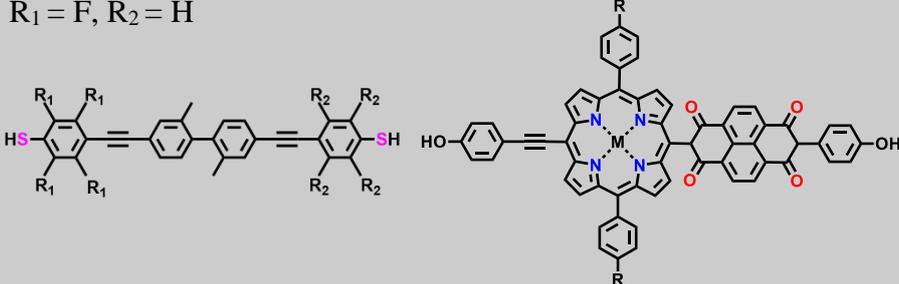 | 27,28 |
| **Non-D-A Type** | 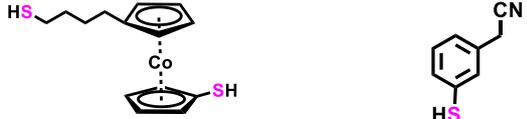 | 29,30 |
| **D-A π Stacked Type** | 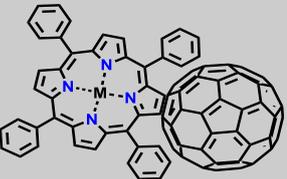 | 31 |

The first column of Table 1 describes various types of molecular design, while representative examples of specific molecular structures and the respective references are presented in the second and third columns, respectively. The first type of molecules is D-σ-A type proposed by Aviram and Ratner (AR model)[23]. Here, tetracyanoquinodimethane (TCNQ) moiety acts as the acceptor (A), bicyclo[2.2.2]octane as an insulator (σ-bridge), and thiofulvalene (TTF) as the donor (D). Similar kind of design was later reported by Perrin *et al.* using phenyl-ethynyl-phenyl as donor (D) and fluorophenyl-ethynyl phenyl as acceptor (A) linked by an insulating ethane moiety[24]. D-A is the next type of molecular design which is different from the AR model; here donor and acceptor moieties are connected directly without a σ-bridge. An example of the respective design, introduced by Yu *et al.*, features an electron-donating bithiophene unit and an electron-accepting bithiazole block[25]. Another example of the D-A design, provided by Diez-Perez *et al.*, involved dipyrimidinyl-diphenyl as diblock, decorated with the thiol anchoring groups to self-assemble on gold electrodes[26]. Elbing *et al.* proposed D-perpendicular-A type molecular design, in which phenyl-ethynyl-phenyl and fluorophenyl-ethynyl phenyl act as donor (D) and acceptor (A) π-systems, respectively, fused by a biphenylic C–C bond[27]. Later, Handayani *et al.* designed a D-perpendicular-A-type complex, in which naphthalene (A) moiety is connected to porphyrin (D) at *meso* position[28]. A further design concept is non-D-A type proposed by Liu *et al.* and featuring cobaltocene asymmetrically decorated with two terminal thiol groups, resulting in an asymmetric electrical potential drop along the molecule[29]. Another example of such a design, reported by Troisi *et al.*, featured 2-(3-mercaptophenyl)acetonitrile with conformational changes upon applied bias producing rectification.[30] Alternatively, D-A π-stacked type design was suggested. A representative example, proposed by Fomine *et al.*, is shown in the Table; it features metalloporphyrin (D) linked to fullerene (A)[31].

Fundamentally, a molecular rectifier must include an asymmetric feature along the direction of charge transport[32–35]. The biggest hurdle in assessing whether the observed rectification behavior is truly molecular or not is the fact that almost any asymmetry in a molecular junction (electrode/molecule/electrode) can generate rectification. An electrode/molecule interface can be ohmic if it does not have notable energy (Schottky) barrier at the interface, provided there is a good matching of the electronic wave functions of the molecule and the electrode[36]. Primarily, the rectification behavior associated with the Schottky barrier originates from the work function mismatch between the contacts to the electrodes rather than from the molecules themselves. However, frequently, naturally grown oxides at the electrode/molecule interfaces generate a noticeable Schottky barrier, leading to an asymmetric voltage profile (potential drop) across the interface that induces asymmetry in the measured current-voltage (I–V) characteristics. Accordingly, such a barrier can likely be the reason behind many observations of small asymmetries in the I–V characteristics of molecular junctions.

Another possible configuration that can lead to rectification in a molecular junction is an asymmetric coupling of a molecule to both electrodes[37–40]. Such a situation can be achieved by employing asymmetric molecules that feature either an asymmetric molecular backbone or different anchoring groups facing the electrodes[26,32,41–43]. However, this strategy has led to diverse results, with comparably low *RR* for various systems, and rectification has also been demonstrated for symmetric molecules[44,45]. Also, both electrodes contacting the molecules in all likelihood (e.g., the same material) can have different surface features such as surface roughness and morphology, making the binding sites for the anchoring groups of the molecules to vary from device to device, leading to variability[46–48]. To reduce the variability, precise control of the electrode morphology and surface chemistry needs to be attained. In addition, a better understanding of the electronic state coupling at these interfaces, charge redistribution upon the junction formation, and charge carrier-injection dynamics is required.

Despite the vast amount of experimental data collected with different experimental platforms, the overall goal of a practical molecular rectifier remains still elusive. A real-life application will require high-quality molecular rectifiers with sharp voltage thresholds, large *RR*, smaller time constants, fast switching speed, retention time, and so on. The major factors that hindered initially the practical application of molecular rectifiers were the lower *RR* values ($<10^2$) compared with commercial CMOS counterparts ($10^5$–$10^8$), limited stability of molecular junctions, and the need for high operating voltages[26,32,42,49,50]. However, during the past five years, a variety of stable molecular junctions with high *RR*s have been reported[51,52, 24,53,54], motivating scientists to work towards commercial implementation of molecular rectifiers. In this review, we showcase recent advances in the field of molecular rectifiers within the different material platforms, including single molecules, self-assembled monolayers, molecular multilayers, heterostructures, and metal-organic frameworks and coordination polymers, and critically analyze the advantages and disadvantages of these systems as well as their potential towards commercialization.

## 2. Single-molecule-based rectifiers

Among various platforms used for the study and realization of molecular rectification, single-molecule rectifiers are particularly important since they materialize the very idea of the truly molecular device and predominantly rely on molecules' intrinsic quantum mechanical properties and hence can exhibit electronic characteristics that cannot be achieved otherwise. It has been generally expected that a molecular rectifier should be longer than one nm in length; otherwise, the tunneling leakage current would overwhelm the function of the molecular component itself. Nevertheless, contacting a single molecule in a device is not an easy task, and to do this in a controlled and reproducible fashion is even more difficult. However, the advances made in the field of manipulation techniques at the nanometer scale, such as mechanically controlled break junctions (MCBJs)[55–57], scanning tunneling microscopy (STM)[58,59], and conductive-probe atomic force microscopy (CP-AFM)[16,60–62], have made it possible to obtain controlled and reproducible results. With improved measurement capability, a variety of new effects beyond the charge transport, including thermoelectricity, quantum interference, and spin transport, were discovered at the single-molecule level[15,63–65]. In the MCBJ technique, by mechanically bending the substrate using a piezoelectrically controlled push-rod, a single atomic contact of a metal is formed using a three-point bending configuration just before breaking the metal wire. Once the wire is broken, a nanogap opens, bridged successively by a molecule introduced from the solution or gas phase, thus forming a single-molecule junction. In contrast, in the STM-BJ technique, molecular junctions are formed in a nanogap made by breaking a metal point contact between the STM tip and a metal substrate. Complementary advances have also been made in first-principle calculation methods. Theoretical approaches based on Green's function theory (many-body theory) have been developed that allow researchers to explore the intrinsic properties of single molecules under non-equilibrium conditions[66,67].

The interpretation of the experimental data in this field is, however, hampered by the difficulty in determining how many molecules are bridged between the electrodes and how many molecules directly contribute to the measured results. Owing to this uncertainty, certain discrepancies emerged not only between experimental data and simulated results but also between datasets collected from different labs. At the same time, techniques such as inelastic electron tunneling spectroscopy (IETS) and surface-enhanced Raman spectroscopy (SERS) at a single molecular level have evolved and profoundly improved our understanding of the chemical and structural behavior of single molecules in the charge transport process. In addition, using a third electrode acting as a gate, allowed tuning the electrostatic potential of the molecule, thereby changing the alignment of the molecular levels responsible for the transport with respect to the Fermi energy of the electrodes, leading to tunable $RR$[68]. The dependence of the $RR$ on the gate voltage increases the confidence that the origin of the observed rectification is truly molecular and not arising from other features of the experimental setup.

Even though rectification in a single-molecule junction was first experimentally realized in 2005, a low $RR$ (<10) and uncertainty regarding the underlying charge transport mechanisms[41,69] have cast doubt on the practical relevance of this concept. However, significant progress has been made during the last five years in this direction, affording single-molecule rectifiers that can potentially rival silicon-based counterparts in their rectification performance. In particular, Aragones et al.[70], using an STM-BJ approach under ambient conditions, reported an $RR$ exceeding $4 \times 10^3$ for 1,8-nonodiyne molecules deposited on low-doped silicon electrodes with high mechanical stability. Besides the efficient rectification achieved in this junction, integrating single molecules with semiconductors added an extra advantage of compatibility with conventional semiconducting circuitry. Alternatively, Perrin et al. demonstrated a gate-tunable, single-molecule diode with an $RR$ as high as $6 \times 10^2$ using an asymmetrically substituted dithiol molecule embedded into a MCBJ setup[71]. The rectification mechanism proposed in that work was directly linked to the molecular structure and attributed to the presence of two conjugated states which were weakly coupled through a saturated linker that broke the conjugation. These findings demonstrate that specific electronic functionality can thus be implemented in single molecules by optimizing the internal molecular structure.

Further in this direction, Schwarz et al. achieved $RR$ values exceeding $10^3$ at bias voltages of less than $\pm 1.0$ V in a two-terminal junction featuring different dithiol molecules with a redox-active metal center[72]. This behavior was mediated by field-induced conductance switching, relying on the redox processes in the metal centers. In a more recent report, relying on the CP-AFM method, Atesci et al.[73] showed that the $RR$ value of Ru-polypyridyl complexes assembled in a monomolecular fashion on a transparent and conducting indium tin oxide (ITO) substrate could be changed by more than three orders of magnitude (between $10^0$ and $10^3$) by simply toggling the relative humidity between 5% and 60% (**FIG. 1**). The key to this behavior is the presence of the two localized molecular orbitals in series, which are nearly degenerate in a dry state but become misaligned under high humidity conditions due to the displacement of counter ions ($PF_6^-$).

Today, the main challenges for single-molecule electronics are the reproducibility of parameters and the improvement of stability of molecular devices. In general, the current detected in single-molecule junctions is very small (~nA), being therefore extremely sensitive to external vibrations, contamination, and environmental conditions, such as humidity and temperature variation. For MCBJ, if the electrode nanogap changes a little, the system will adopt a different configuration that might be either stable or unstable. The main difference between the stable and unstable junctions can then be the difference in the structure of the immediate contact region between the molecule and electrodes, which can vary from junction to junction and heavily influence reproducibility. Furthermore, the reproducibility can be affected by possible molecular degradation, which is a factor that also has to be considered. Moreover, experimental realization of molecular rectification mostly requires a relatively high bias voltage (≥ 1 V), at which the stability of a single molecule junction can become a limiting factor. Another fundamental challenge is the integration capability, i.e., design and fabrication of single-molecule junction platforms capable of serving as complement parts in the existing CMOS circuits. Finally, although most of the relevant molecules can be synthesized in large quantities and cost-effectively, most widely used techniques, such as electron-beam lithography for nanoelectrode fabrication, are still too expensive for large-scale production.

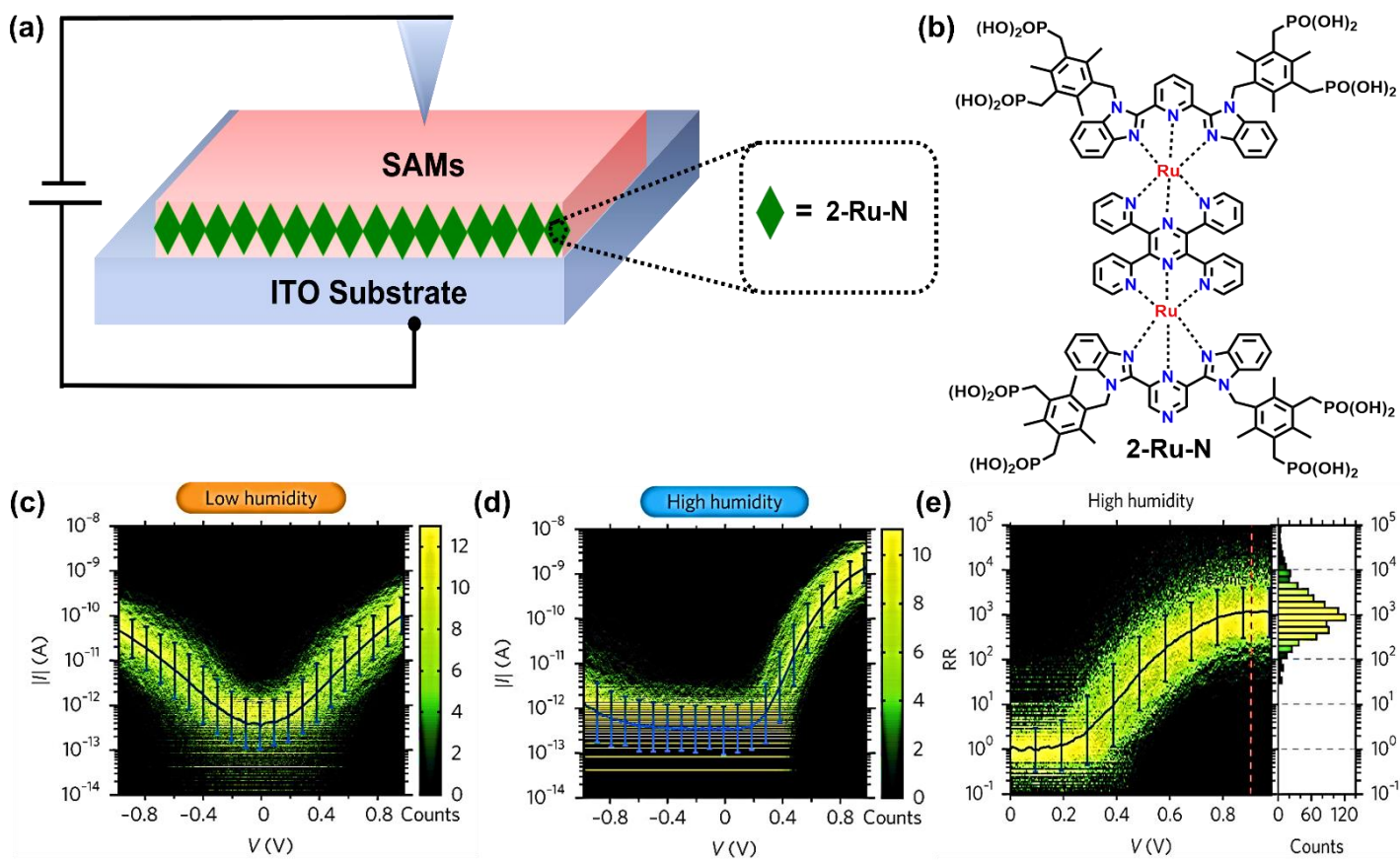

**FIG**. **1 An example of single-molecule rectifier:**[73] **(a)** Schematic of molecular junctions with Ru(II)-polypyridyl molecules assembled on ITO substrate and contacted by CP-AFM probe, **(b)** the chemical structure of the molecule. The counter anions are not shown here for simplicity. **(c, d)** 2D histograms of current-voltage (I-V) curves for low **(c)** and high **(d)** humidity cases. **(e)** 2D histogram of logarithmically binned *RR* for the high humidity case (right panel) and 1D histogram taken at V=0.9 V (left panel). Overlaid in blue is the mean of the Gaussian fits at each bias voltage bin. The error bars represent the half widths of the fits.

## 3. Self-assembled monolayers-based rectifiers

The use of 2D molecular ensembles, such as self-assembled monolayers (SAMs), is more appealing as a working practical device than single molecules due to the ease of fabrication, compatibility with large-scale production, and comparably low costs[74–77]. A variety of methods can generally be used for the assembly of densely packed SAMs on different substrates; interested readers are directed to reviews on monolayer formation in general[19,78,79] and their assembly on specific substrates, such as metals[19,80], oxides[81–83], and oxide-free Si[84–86]. Among different monomolecular systems, SAMs with thiolate anchoring groups, originating mostly from thiols or dithiols, turned out to be especially popular in the context of large-area molecular junctions[87–93]. The advantages of these systems are not only the high affinity of thiols or dithiols to coinage metals but also their relative simplicity, reproducibility, and versatility. The disadvantages include a comparably long self-assembly time, defect formation, and high chemical reactivity which frequently leads to partial oxidation of SAM-forming molecules[19,94].

Generally, the electronic structure of molecules embedded between the two electrodes and work functions of the electrodes define the energy level alignment in a given junction, including possible charge transport barriers. At the same time, upon assembling a junction, charge transfer between the molecules and electrodes usually takes place, with the extent depending on the strength of bonding (weak or strong electronic coupling), until an equilibrium state, associated with possible realignment of the energy levels, is achieved. Consequently, not only the molecules but the entire junction must be taken into account while determining the tunnelling barriers and overall charge transport behavior[95].

With all the existing knowledge gained over the last two decades, researchers can now frequently rationally design and even fine-tune the rectification behavior of a given molecular junction by modifying the chemical structure of SAMs and/or by varying the characteristics of the electrodes. It is important to note that the mechanism of rectification can vary considerably depending on the molecular structure and specific experimental platform used for a particular study. Initially, very low rectification values ($RR$ <30) were reported for molecular ensembles. A breakthrough in this direction was the report of Nijhuis *et al.*, in which SAMs of alkanethiols with a terminal, redox-active ferrocene (Fc) moiety were asymmetrically placed between the bottom Ag electrode and top eutectic GaIn (EGaIn) electrode [95], with the device exhibiting a $RR$ of ~$1.0 \times 10^2$ at ±1 V[96,97,98]. The rectification behavior was associated with the constructive participation of the ferrocene HOMO in the charge transport process at negative bias[97,98]. Using a combination of two ferrocene moieties (Fc−C≡C−Fc) instead of one, the same group later fabricated a molecular diode with a $RR$ of $1.0 \times 10^3$ [34]. The operation of this diode relied on sequential tunneling involving both HOMO and HOMO−1 of ferrocene molecules positioned asymmetrically inside the junction. Further extensions of these activities were controlling the direction of rectification in a molecular diode by the variable placement of ferrocene within the molecular backbone[87] and analyzing the factors which affect the performance of molecular diodes[87,89]. Following a similar experimental approach but using fullerene ($C_{60}$) instead of ferrocene, Qiu *et al.* employed undecanethiolate SAMs with $C_{60}$ tail group in a junction with the bottom Ag and top EGaIn electrodes and reported an $RR$ of up to 940 at ±1 V[99]. The exact $RR$ value varied however to some extent dependent on the fullerene density, the composition of the top contact (EGaIn), and the energy of the lowest unoccupied π-state of the SAMs-forming molecules (**FIG. 2a,b**). Compared to ferrocene, fullerene-based junction showed a higher $RR$ value. It can be ascribed to the lower energy difference in electrode Fermi level and LUMO (~4.5 eV) of the fullerene, besides the larger surface, spherical symmetry of $C_{60}$ molecules, and asymmetric metal (contact)-fullerene interactions. The junctions also show hysteresis in the I-V curve upon applying positive bias above 0.45 V, which further increases at +1.0 V. The authors have not shed light on a reason for the hysteresis, but one can speculate that it is triggered by the involvement of ambient moisture/water vapor, possible under the ambient conditions of the experiment. This underlines the importance of measuring *RR* under entirely controlled conditions to improve the reproducibility of the results and to avoid their distortion by external factors.

The next major step was the use of Pt substrate instead of Ag, resulting in an *RR* exceeding $10^5$ (at a bias of ~3.0 V), comparable to CMOS-based devices ($10^5$-$10^8$) - for the molecules with the Fc–C$\equiv$C–Fc termini[100]. A noticeably lower but still reasonable *RR* of $2.5 \times 10^4$ and a resistive ON/OFF ratio of $6.7 \times 10^3$ were reported for a molecular junction featuring the Ag and EGaIn electrodes and SAM of $S(CH_2)_{11}MV_2^+ X^-_2$ (where MV is methyl viologen and $X^-$ is the counter ion, **FIG. 2c,d**),[53] which is a promising outcome. Further significant advancements include a transition from direct to inverted charge transport Marcus regions in molecular junctions via molecular orbital gating[101] and reversal of the direction of rectification induced by Fermi level pinning at molecule-electrode interfaces in redox-active tunneling junctions[102]. Most recently, SAMs of $S(CH_2)_{11}BIPY$ (BIPY= bipyridine) complexed with a metal ion (cobalt or copper) have been integrated into molecular junctions with Au and EGaIn electrodes[103]. The junction featuring BIPY-CoCl$_2$ moiety showed a reasonable *RR* value of ~82 at ±1.0 V due to the constructive involvement of the dominant molecular orbital, such as the LUMO, at positive bias only. Rectification behavior disappeared, however, for BIPY-CuCl$_2$ because of the symmetrical involvement of the molecular orbitals (HOMO or LUMO) at both positive and negative bias.

Another important advancement was the use of interstitially mixed monolayers, which are an attractive alternative to the traditional single-component and mixed SAMs. In interstitially mixed monolayers, small molecules fill the interstices between the bulky molecules via repeated molecular exchange minimizing the defects within the monolayer. The use of such specifically designed films can potentially enhance the stability and reproducibility of molecular junctions without degrading their function and reliability. For instance, Kong *et al.* fabricated interstitially mixed SAMs using *n*-undecanethiol terminated with 2,2′-bipyridyl (HSC$_{11}$BIPY) for matrix and *n*-octanethiol (HSC$_8$) as reinforcement material[104] with a continuous exchange of molecules in the SAMs occurring during rectification. The experimental results combined with theoretical simulations confirmed a defect-free monolayer structure, yielding high operation stability in a noticeably broader voltage range (± 3.3 V) compared to traditional mixed SAMs.

Other potential candidates for molecular rectification are SAM-forming biomolecules, such as oligopeptides, proteins[105], and DNA[106–113]. A particular advantage of DNA is the possibility of fine-tuning its charge transport properties by subtle alteration of its conformation via the ionic environment, methylation, or intercalation with metal ions and complexes[114–116]. In particular, in one of representative studies, a DNA-based molecular rectifier was constructed by site-specific intercalation of small molecules, such as coralyne, into an 11 base-pair DNA-complex, providing an *RR* value of ~15 at ±1.1 V[117]. Also, an oligopeptide-based diode with platinum electrodes featuring peptide chains comprised of two pairs of amino acids, alanine–lysine and threonine–alanine, has been reported[118]. The length of the peptide chains and the arrangement of the amino acids played a pivotal role in its performance.

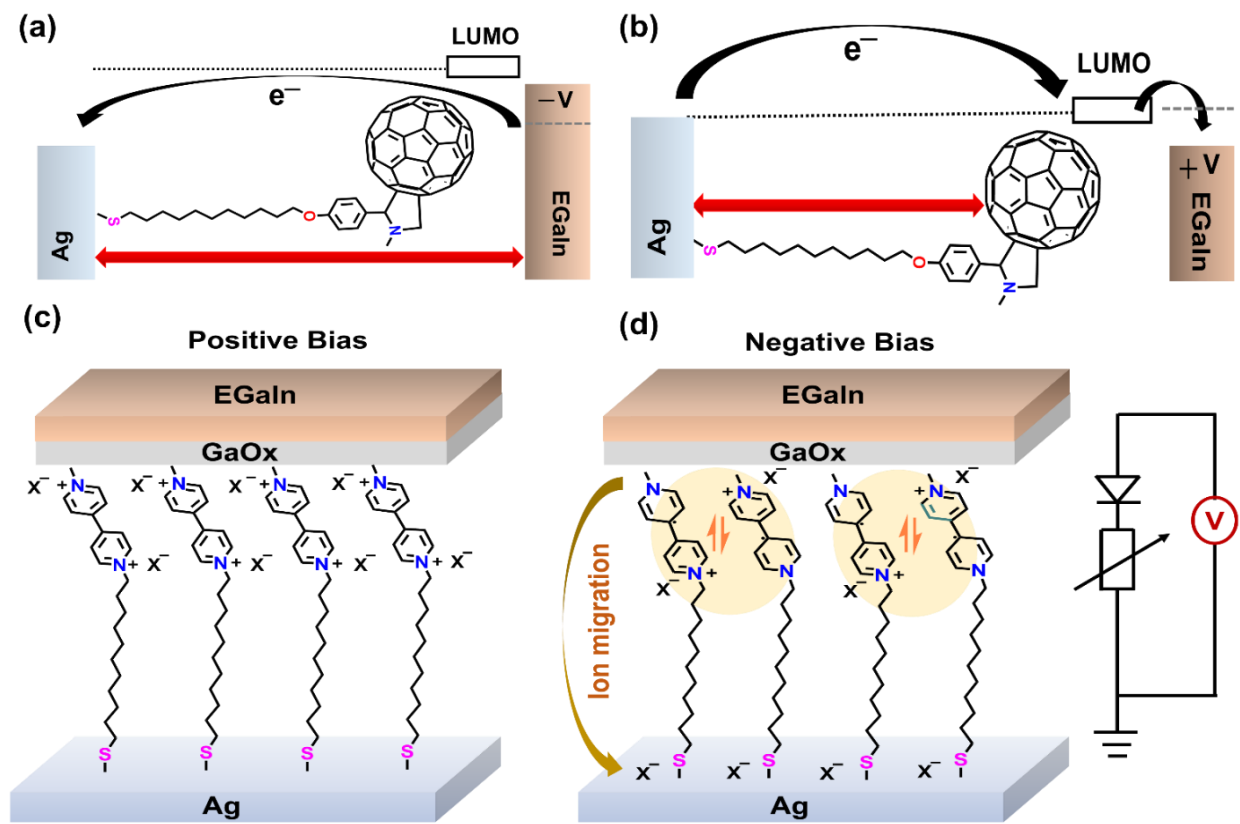

**FIG. 2 Examples of SAM-based molecular rectifiers:** (a,b) Schematic of rectification recorded in undecanethiolate SAMs with fullerene ($C_{60}$) tail group embedded in the junction with bottom Ag and top EGaIn electrodes. At negative bias (a), LUMO is outside the conduction window (the region between the Fermi levels of Ag and EGaIn), whereas at positive bias, (b) LUMO is brought into the window; as a result, electrons tunnel to the LUMO and then hop to the EGaIn electrode. (c,d) Schematic illustration of an electric field-driven, dual-functional molecular device featuring alkanethiolate SAMs terminated with methyl viologen moiety at a positive (c) and negative bias (d).

### 4. Multilayer molecular rectifiers

Over the years, multilayer structures in nature (butterfly wings, birds feather, etc.) have strongly motivated researchers to design and develop a wide range of artificial multilayer-based coatings and devices with established superior properties in the fields of energy, metamaterials, optoelectronics, power and flexible electronics, signal systems, etc[119]. Multilayer-based molecular rectifiers are not an exception in this context, gaining significant attention due to their high internal quantum efficiency, potentially high *RR* and transconductance, and a feasible way of junction fabrication. In this aspect, molecular bilayers employing the donor (D)-acceptor (A) concept, mentioned in section 1 and illustrated to some extent in Table 1, are most prevalent. In general, under forward bias, electrons flow from the acceptor to the donor side (AR model[23], see **FIG. 7**) provided the coupling between the donor and acceptor is stronger than that to the electrodes, which, however, should still be sufficiently good for acquiring high currents and getting rectification behavior[120–122]. Conversely, if the coupling between the donor and acceptor layers is weaker than that to the electrodes, electrons flow from donor to acceptor at a forward bias (anti-AR model, see **FIG. 7**)[123]. This D-A concept was in particular tested in bilayers featuring $C_{60}$ and pentacene (Pn) films deposited on Cu(111) and contacted from the top by an STM tip[124]. The resulting assembly epitomized an AR-type rectifier with *RR* values close to $10^3$ but also showed a lower *RR* value near the edges of the bilayers as a consequence of confinement and tip-induced Stark effects. Apart from the molecular structure of a bilayer, the extent of rectification also depends on a variety of parameters, such as the nature of the contacts, internal dipoles, charge polarization, presence of trapped molecules, and details of the bilayer preparation (including solvents, temperatures, and operating voltage) experienced by a particular system[125,51]. In particular, Lopes *et al.* have reported a temperature-regulated inversion of rectification direction (IRD) in molecular diodes employing copper phthalocyanine (CuPc) and hexadecafluorophthalocyanine ($F_{16}$CuPc) bilayer[126]. An *RR* of $10^2$ with I(−V) higher than I(+V) has been observed at temperatures below 50 K, but decreased progressively with the increasing temperature. At T = 220 K, I(+V) and I(−V) were equal, while at room temperature I(+V) exceeds I(−V), thus displaying temperature-driven IRD behavior. The IRD phenomenon was explained on the basis of trap-mediated Poole-Frenkel mechanism and tip-induced Stark effect. Other AR rectifiers, featuring D-A layers with comparably large HOMO-LUMO gap, were reported to rectify at both low and high temperatures with an *RR* of ~300[10].

It is also important to have a wide frequency response in molecular junctions, which has not been achieved until recently, being limited by low frequencies. The frequency range can, however, be significantly extended by constructing an electronic device utilizing organic bilayer heterostructures. For instance, junctions comprising ultrathin $F_{16}$CoPc/CuPc bilayer (1 nm/7 nm) exhibit rectification behavior at a frequency of up to 10 MHz[127]. The fabricated devices showed an *RR* of ~400, ascribed to the different interfacial properties of the top and bottom Au electrodes. The high-frequency response was achieved due to efficient charge injection and transport from the bottom Au contact to phthalocyanine. In another study, molecular junctions featuring Fe(II) bis-terpyridine-appended dithienylethene (Fe-tpy-DTE-tpy)$_n$ photochromic film of variable thickness, fabricated by layer-by-layer (LbL) technique, were studied (**FIG 3a**)[128]. The junctions displayed a reasonable *RR* of 200 in both photostationary states along with an ON/OFF ratio of 120.

Besides the temperature and light factors, the effect of solvent on the rectification behavior of bilayer and multilayer molecular junctions was studied as well[129]. In a related study, fluorene (FL) and benzoic acid (BA) were used to form a bilayer where $H^+$ ions were replaced by lithium ions forming lithium benzoate (LiBA) and providing carbon/FL-4.5nm/LiBA-4.0nm/carbon stacking configuration (**FIG 3b**)[129]. The conductance of the respective junctions, containing mobile ions, decreases drastically at the presence of a polar solvent, triggered by the electric-field-induced solvent and ions reorganization. Whereas *RR* of these junctions was initially very low (1.3 at ±1V), it increased, however, to 18 after repetitive I-V scanning, and the junctions could be reset at 0 V. In another study, McCreery and co-workers have fabricated both single and bilayer molecular junctions consisting of anthraquinone (AQ), bis-thienylbenzene (BTB), and AQ-BTB combination (**FIG 3c**). The polarity of photoconduction and rectification direction of the respective devices depended strongly on the electronic structure of the involved molecular layers[130]. Further, the same group has synthesized a multifunctional trilayer by stacking AQ, BTB, and LiBA layers and integrating this assembly into a junction with carbon electrodes as bottom and top contacts (**FIG 3d**). The single AQ layer, AQ/BTB bilayer, and LiBA/BTB/AQ trilayer assembly have been checked for the I-V performance, exhibiting unique characteristics, including *RR* of 28 and 75 for the bilayer and trilayer, respectively[131].

Several research groups have shown that multilayer molecular rectifiers can also be used for brain-inspired memory devices, which display exceptional computing performance, outstanding tunability, and excellent adaptability[132,133,134]. In particular, a combination of $C_8$-BTBT bilayer with a modulated Schottky barrier diode was used to fabricate phototunable synaptic-like devices for neuromorphic computation to overcome the von Neumann bottleneck. The device featured Au/C8-BTBT/Au configuration grown on Si/SiO$_2$ substrate, with the thickness of the active layer and channel length being set at 3 nm and 40 µm, respectively. Under a small applied bias, the fabricated devices displayed an exceptional *RR* of ~$10^5$ with an energy consumption of 13.6 pJ[135], along with the striking light switching performance ($I_{light}/I_{dark}$ =100 at 2 V), excellent state holding ability, and variable synaptic behavior. These exceptional properties make the respective molecular devices potential candidates for energy-efficient organic optoelectronic computing elements with nearly ideal characteristics.

Generally, multilayer-based molecular rectifiers have numerous advantages, making them potentially eligible for a variety of advanced technological applications. However, they also possess several serious drawbacks, such as complicated fabrication processes, limited availability, high production costs, and complex energy band arrangements, which should be overcome or at least diminished for the practical realization and commercialization of the respective devices.

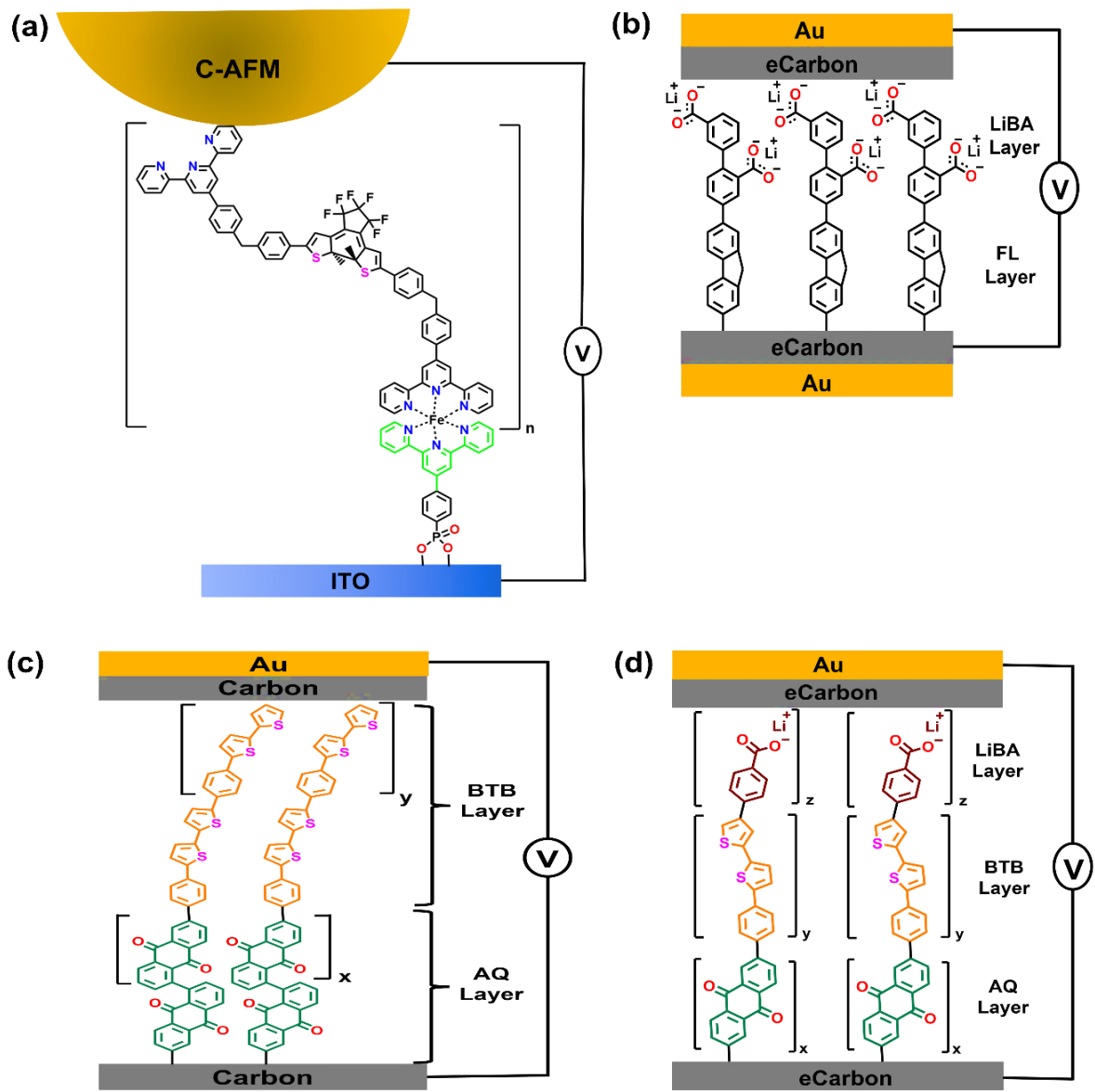

**FIG. 3 Schematic illustration of molecular junctions featuring** (a) (Fe-tpy-DTE-tpy)$_n$ multilayer assembled via LbL technique; (b) FL$_{4.5}$/LiBA$_4$ bilayer; (c) BTB/AQ bilayer; and (d) a LiBA/BTB/AQ trilayer. The molecules were assembled on either ITO (a) or eCarbon/Au (b-d) substrates; a CP-AFM tip (a) or eCarbon/Au (b-c) were used as the top electrode.

## 5. Heterostructure-based rectifiers

Remarkable progress has been made in designing and synthesizing conjugated molecules. However, harnessing useful work from these molecules to perform as a rectifier is a more daunting task. In this aspect, non-functionalized molecules, including conjugating ones, which provide symmetric tunneling charge transport can, at first sight, hardly perform as molecular diodes[136–138]. Nevertheless, these molecules are able to demonstrate desired, asymmetric I-V characteristics in heterojunction structures as their interfacial band alignment can be altered via combining with other moieties. In this regard, a variety of design concepts for the fabrication of hybrid molecular junctions or heterojunctions has been developed, where inorganic materials, such as nanoparticles and 2D semiconductors, were combined with molecular films[139–141]. For instance, in one of the relevant studies, ZnO nanoparticles were electrostatically assembled over organic molecular films such as rose Bengal (RB) and copper phthalocyanine (CuPc) to fabricate hybrid molecular junctions (n-type Si/n-ZnO or p-ZnO/organic molecules/Hg or n-type Si/organic molecules/n-ZnO or p-ZnO/Hg)[142]. Due to the donor-acceptor character of the nanoparticles and organic molecules, respectively, charge flow was feasible only in one direction, resulting in current rectification. In another study, metal-phthalocyanines and transition metal dichalcogenides ($MoS_2$ and $WSe_2$) were used as organic and inorganic materials to fabricate a heterojunction (by an electrostatic adsorption process) featuring current rectification (**FIG 4a**)[143]. Here, the central metal atom played a crucial role; depending on the character of its 3d band the strength of the magnetization vector changed, triggering alignment of planar metal phthalocyanine molecules on the electrode surface. Such an alignment altered the molecular orbitals and thereby affected the energy levels at the interface, which overall influenced the *RR* values. In another representative study, $MoS_2$ and $WSe_2$ layers have been combined with thiolate SAMs on Au as organic-inorganic heterojunctions (**FIG 4b**)[144]. The number of the $MoS_2$ and $WSe_2$ layers and the length of the molecular chain of the SAM-forming molecules were varied. In the case of the SAMs only, no asymmetry was observed (an *RR* of ~1); however, the I-V curves changed dramatically with the attachment of $MoS_2$ and $WSe_2$, becoming asymmetric and yielding *RRs* of $1.79 \times 10^3$ and 2.31 at $\pm 1$ V for the $OPT2/MoS_2$ and $OPT2/WSe_2$ junctions (OPT2 = biphenyl-4-thiol), respectively.

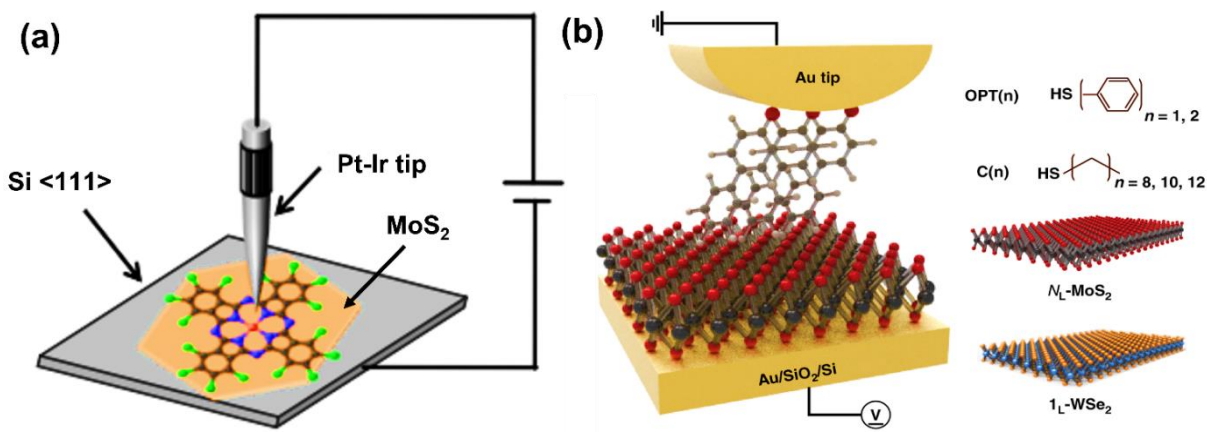

**FIG. 4 Schematic illustration of molecular junctions featuring** (a) $MoS_2$/MPc (metal-phthalocyanines), and (b) $MoS_2$ or $WSe_2$ /SAMs heterostructures for studying nanoscale rectification. Figures (a) and (b) are reproduced with permission from ref.[143] and ref.[144], respectively.

## 6. Coordination polymers/MOFs-based rectifiers

Presently, research areas related to metal-organic frameworks (MOFs) and coordination polymers (CPs) formed by coordinating metal nodes and organic linkers are rapidly flourishing due to the tunable electronic and optoelectronic properties of these systems[145–147]. However, so far, MOFs have been hardly explored in molecular electronics because of the insulating nature of organic linkers and a poor overlap between the π-orbitals of these linkers and the d-orbitals of the metal centers, resulting in electrical conductivity of $10^{-10}$ S/cm or even lower. The conductivity can, however, be noticeably improved by different means, such as coupling the metal sites to the redox-active linkers, doping with redox-active guest molecules, etc.[121,148,149,150,151] Besides these options, the introduction of noncovalent interactions within coordination complexes, such as H-bonding, π-π interactions, van der Waals forces, anion-π interactions, etc., also leads to enhancement in the MOF conductivity as described in numerous reports, reviews, and perspectives[10,120,152,153].

Even though polymers are generally regarded as semiconducting, their limited thermal stability makes them unsuitable for molecular electronics. In this context, MOFs could probably be more suitable forthcoming applicants in molecular electronic devices, providing us a new domain, 'MOFtronics'[154]. From this perspective, we mostly emphasize here the use of MOFs or CPs as rectification devices, linking the rectification behavior to their structural properties. In 2017, first efficient, CP-based molecular diodes have been developed using Cd(II) coordination polymers, such as [Cd(4-bpd)(SCN)$_2$]$_n$ and [Cd$_4$L$_2$(NCO)$_6$]$_n$ complexes, where 4-bpd is 1,4-bis(4-pyridyl)-2,3-diazo-1,3-butadiene and L is (E)-2-methoxy-6-((quinolin-8-ylimino)methyl)phenol (**FIG 5a**)[155,156]. Upon exposure to light, these devices displayed distinct rectification behavior, which was further optimized by the structural variation of the CP linker moieties.

Before 2019, the overall reported *RR* values for the best performing MOF/CP-based diodes were in the range of $10^2$-$10^3$, which is noticeably lower compared to analogous CMOS-based devices ($10^5$-$10^8$). However, a recent report by Ballav *et al.* demonstrated a much higher *RR* value of $10^5$ for a device featuring Cu(II)-BTEC CPs, where BTEC is 1,2,4,5-benzenetetracarboxylic acid[157]. To enhance the conductivity of CPs, TCNQ (7,7,8,8-tetracyanoquinodimethane) was doped into the Cu-BTEC thin film. Such doping created electronically coupled conducting channels via Cu$^{2+}$····TCNQ coordination linking, leading to an enhancement in conductivity by a factor of up to $7.8 \times 10^6$ compared to the non-doped CP film. The high *RR* value was attributed to the resemblance of the Cu(II)-BTEC film structure with a regular p-n junction, where half of the film was conductive (due to the doping) while the other half was insulating. Apart from this configuration, certain counter anionic species have been observed as another factor affecting the electrical conductivity of MOFs. In particular, a recent study has reported that BF$_4^-$ species were used as guests in a three-dimensional network of Zn(II) MOFs with neutral bispyrazole based ligands[158]. In these systems, the porous channels formed inside the MOFs were populated with various H-bonded anions and anion-solvent clusters, which resulted in an enhanced electrical conductivity. When integrated into a two-terminal junction configuration with the bottom ITO and top Al electrodes, they provided current rectification, even though with quite low *RRs* of 15-19 only (**FIG 5b**). In a further study, electronically conducting MOFs (EC-MOFs) have been employed as an effective interlayer material to modulate the Schottky barrier ($\Phi_B$) in the self-powered Schottky diodes, which is essential for photodetector application as well. A series of highly oriented triphenylene-based EC-MOF thin films, M$_3$(C$_{18}$H$_6$X$_6$)$_2$ (M = Ni or Cu; X = O or NH), has been utilized to optimize the silicon-based Ag/EC-MOF/n-Si Schottky junctions (**FIG 5c**)[159]. $\Phi_B$ was successfully tuned from 0.67 eV in Ag/n-Si to 1.11 eV in Ag/EC-MOF/n-Si junction via varying EC-MOF components. However, preparing such high-quality EC-MOF thin films is still a daunting task.

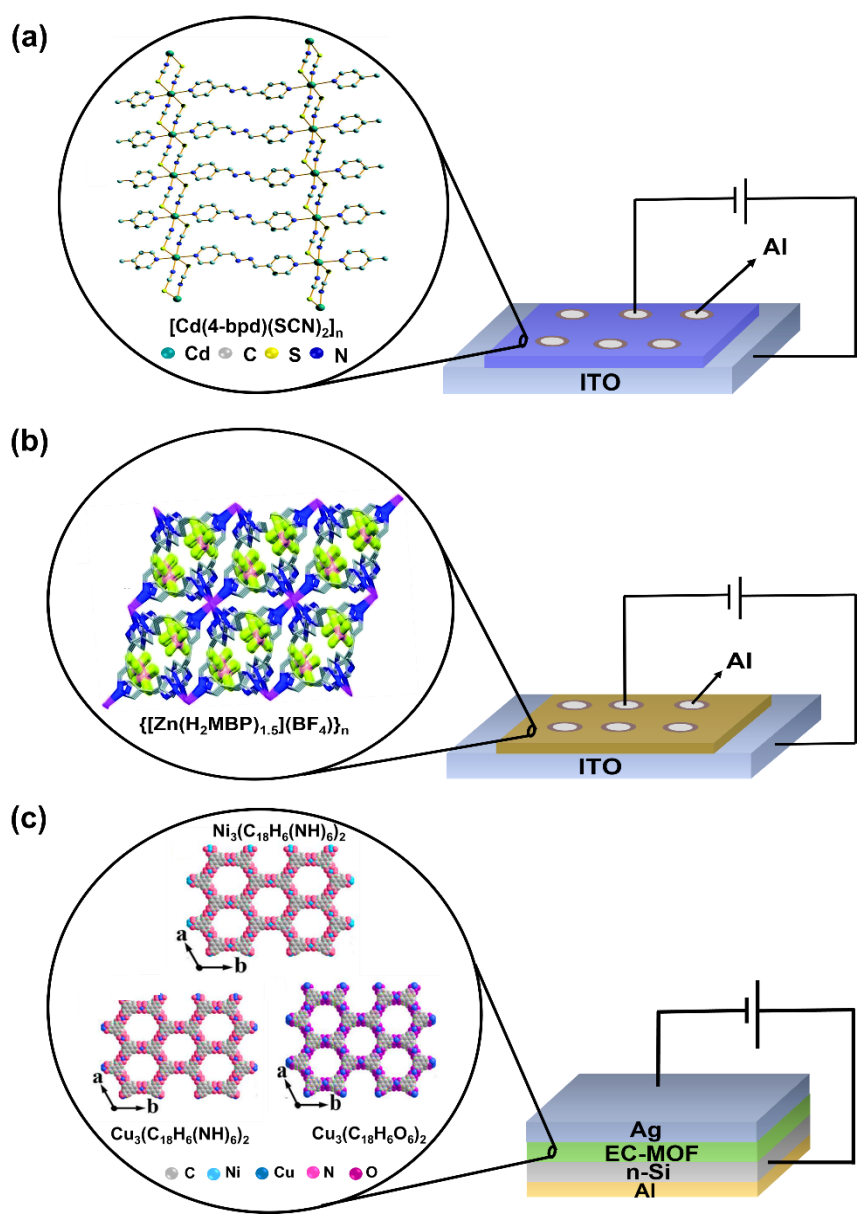

**FIG. 5 MOFs-based molecular junctions exhibiting current rectification characteristics, with the structure of the sandwiched MOFs additionally shown.** (a,b) Schematic illustrations of (a) [Cd(4-bpd)(SCN)$_2$]$_n$[160] and (b) {[Zn(H$_2$MBP)$_{1.5}$](BF$_4$)}$_n$[161] based molecular diodes employing ITO and Al as bottom and top contacts. (c) Schematic illustration of Ni$_3$(C$_{18}$H$_6$(NH)$_6$)$_2$, Cu$_3$(C$_{18}$H$_6$(NH)$_6$)$_2$, and Cu$_3$(C$_{18}$H$_6$O$_6$)$_2$ based diodes employing n-type Si and Ag as bottom and top contacts[159]. The MOF structures in (a), (b), and (c) are reproduced from refs.[160,161,159] with permission.

Recently, the basic molecules of crystalline organic semiconductors, such as anthracene and fullerene, have also been applied as linker moieties to fabricate surface-anchored MOFs (SURMOFs)[162]. By using a LbL approach, crystalline Cu$_2$(adc)$_2$(dabco) [adc = 9,10-anthracene dicarboxylate] and C$_{60}$@Cu$_2$(bdc)$_2$(dabco) [bdc = 1,4-benzene dicarboxylate, dabco = 1,4-diazabicyclo[2.2.2]octane] layers, denoted as p-SURMOF and n-SURMOF, respectively, were successively grown on a pre-functionalized gold electrode forming a p-SURMOF/n-SURMOF assembly (**FIG 6a,b**). The MOFs of different conductivity were thus combined over a crystalline heterointerface, building the molecular devices exhibiting a diode-like behavior with an *RR* of approximately six orders of magnitude. **FIG 6c** shows the junction's energy profile diagram, where the purple and green lines represent the HOMO and LUMO of the p/n-SURMOFs. Here, anthracene and fullerene are the active moieties, mediating the charge transport in the assembly via a charge hopping mechanism, which is quite common in organic semiconductors at room temperature.

Generally, the p-n junction configurations obtained for the exemplary SURMOFs by either selective doping[157] or successive growth[162] can be extended to other MOF systems, enabling to fabricate p-n-like heterojunctions with exceptionally high *RR* values. This underlines the potential of MOFs and CPs in the context of molecular rectification. However, despite the progress achieved so far, large-scale production of reliable MOF-based devices is still difficult as preparation of high-quality and well-ordered crystalline SURMOFs is yet a challenging task. Also, ligand design and coordinating chemistry are somewhat limited and hence require new ideas and concepts for further development. It is nevertheless believed that the scalable formation of conducting MOFs via controlled preparation routes and their facile fabrication on conducting substrates will make MOFtronics renowned in the near future.

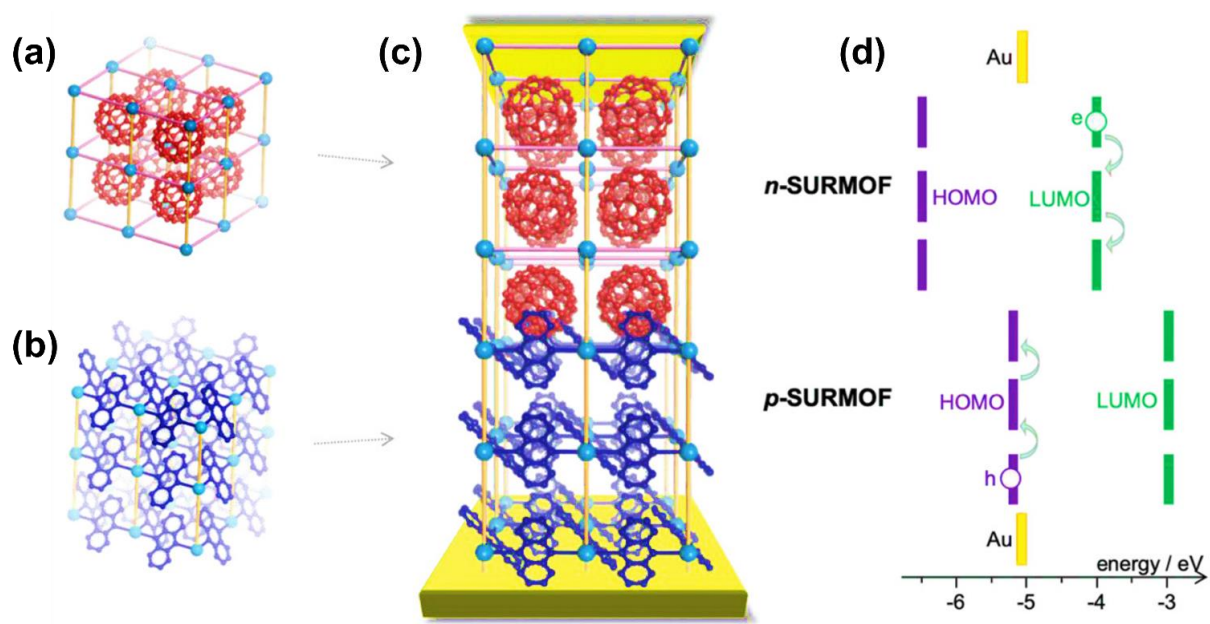

**FIG. 6 An example of SURMOF-based rectifier:** Schematic illustration of (a) fullerene- ($C_{60}$) and (b) anthracene-containing n-SURMOF and p-SURMOF. (c) Schematic illustration of a junction featuring a p/n-SURMOF bilayer with crystalline heterointerface and top and bottom Au electrodes. (d) Energy profile diagram of the bilayer junction with the HOMO and LUMO of the p/n-SURMOFs and schematic illustration of the charge transport mechanism. The figure is reproduced from ref.[162] with permission.

## 7. Plausible mechanisms of rectification

Along with the description of different molecular rectifiers, the plausible mechanisms that govern their rectification behavior should be shortly discussed. The basic theoretical models are graphically represented in **FIG 7**. In the sequential tunneling model proposed by Aviram and Ratner[23] (**FIG. 7a**), individual tunneling steps, with the rates denoted as $\Gamma_n$, occur through the molecular energy levels of the acceptor (A) and donor (D) units, with three tunneling barriers being involved. The respective units, separated by an electrical spacer (σ bridge), respond differently to the applied bias, resulting in their alignment or misalignment depending on the bias sign, emphasized by a difference in the onset of the resonant tunneling for the two bias directions. However, frequently, the A and D units belong to the same molecule and are not electronically decoupled, so that the intermediate tunneling step can be neglected, with only two steps left. This assumption is a basis for the Kornilovitch-Bratkovsky-Williams (KBW) and Datta-Paulsson (DP) models of molecular rectification, illustrated in **FIG. 7b** and **FIG. 7c**, respectively. A further key assumption of these models is the different couplings of the molecular electronic system with those of the electrodes, resulting in asymmetric level shifting depending on the bias sign. The mechanism of the asymmetry is, however, different in both models. In the KBW model, the rectification is mediated by asymmetric tunneling barriers, with the level shifting by an electric field. This leads to a shift in onset voltage for resonant tunneling depending on bias direction. In contrast, in the Datta-Paulsson (DP) model, a different differential conductance in the region of resonant tunneling plays a major role, mediated by asymmetric charging[163] (**FIG. 7c**). The higher the charging energy of a given molecular state, the smaller is the differential conductance. The KBW and DP models can also be combined, resulting in a generalized description of molecular diodes (**FIG. 7d**), especially useful for asymmetric coupling of the energy levels typical of asymmetric molecules. Within this combined model, the discrete energy levels can be shifted due to electric field, charging, or a combination of both these factors.

It should be noted that it is frequently difficult to distinguish between the three discussed mechanisms (Aviram-Ratner, asymmetric field, asymmetric charging) in a particular experiment. For instance, a strict Aviram-Ratner diode is difficult to realize practically since a complete decoupling of the donor and acceptor energy levels is not always guaranteed[164]. Moreover, charge transport experiments frequently suffer from an uncontrolled coupling between the terminal groups of the molecules and the electrodes due to a not entirely controlled character of the terminal group-electrode interaction and random contact geometries[165].

It should also be noted that the described basic theoretical models do not exclude further system-specific models, reflecting specific features and parameters of molecular systems, such as involvement of several energy levels, redox processes, quantum interference, change of molecular conformation, etc. Such models, along with quantum mechanical simulations, are frequently derived in context of specific experiments to rationalize the results.

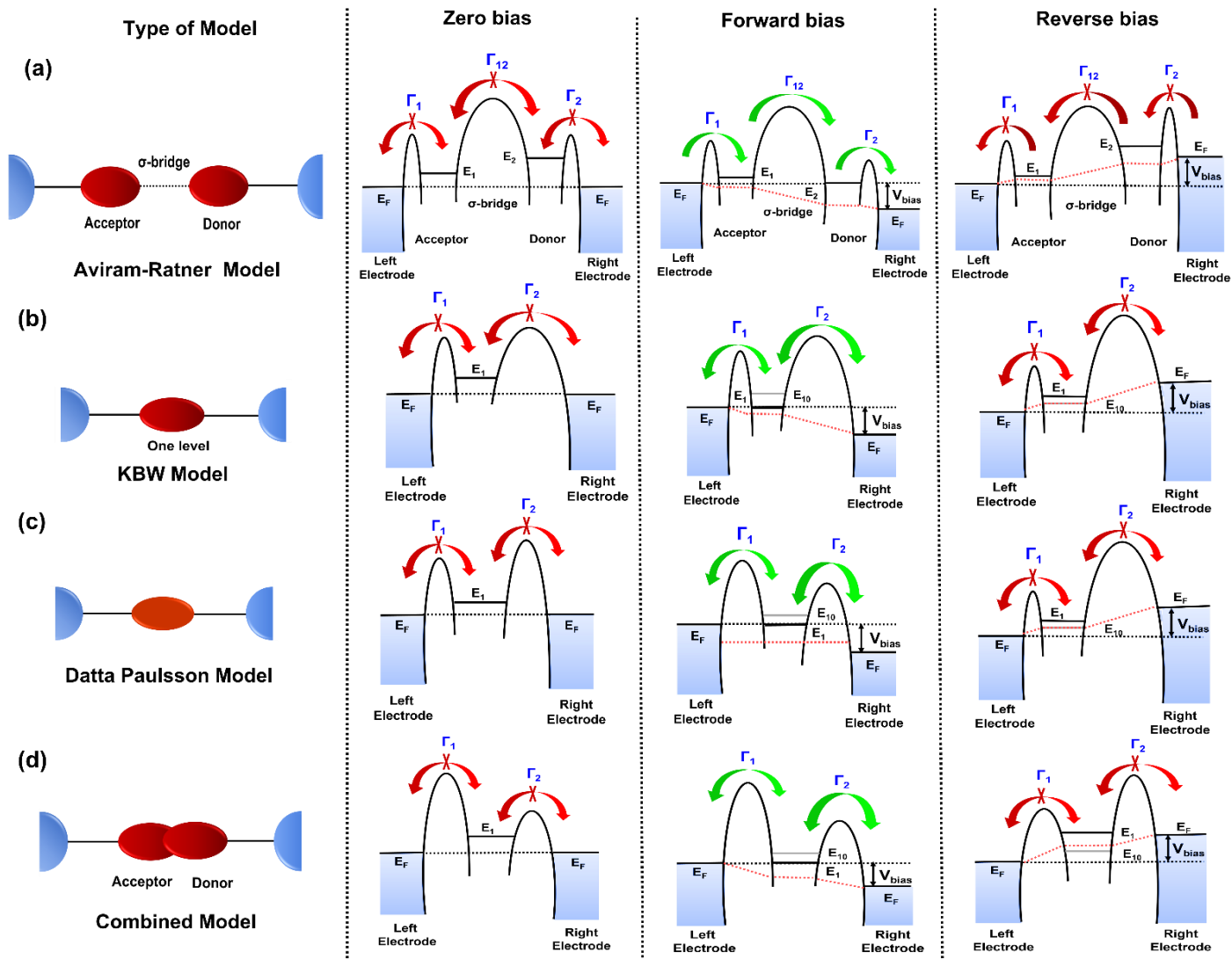

**FIG. 7 Theoretical models employed for explanation of current rectification in molecular junctions.** (a) In the A-R model, the energy levels of the donor and acceptor moieties, separated by an electrical spacer, respond differently to the applied bias, resulting in their alignment or misalignment depending on the sign of the bias. (b) In the KBW model, the couplings of the molecular energy level to the electronic systems of both electrodes are different. Accordingly, the energy level shifts asymmetrically by an electric field, leading to rectification. (c) In the DP model, the diode characteristics originate from asymmetric energy level shifting through the different charging energy. (d) In the combined model, asymmetric shifting of the energy level is mediated both by an electric field and charging energy. Only one energy level is used in this model for the sake of clarity. $\Gamma_n$ represents the rates of individual tunneling steps. The figure is adapted from ref.[166] with permission.

## 8. Comparison of different systems and related implications

Discussing the different concepts of molecular rectification and providing representative examples in this context, we think that it can be useful to present a short overview of the discussed systems to compare their performance, represented mostly by the *RR*, with each other and the conventional CMOS-based devices. Such an overview is shown in Table 2. Whereas the *RR* values differ largely between the different molecular systems prepared within the different design concepts, in many cases, they are sufficiently high to compete with the CMOS-based devices. Indeed, the SAM, multilayer, and CP/MOF approaches deliver $RR$ values on the order of $10^5$ and even higher, with the ultimate values of $6.3 \times 10^5$ and $\sim 10^6$ for the SAM of Fc–C≡C–Fc-decorated alkanethiolates on Pt and $Cu_2(adc)_2(dabco)/C_{60}@Cu_2(bdc)_2(dabco)$ SURMOF assembly, respectively. Thus, the value of $RR$ is no longer the critical issue for molecular rectifiers, even though its further increase is, of course, highly desirable. The major issues in the context of application and commercialization of molecular rectifiers are reliable and large-scale-production-adaptive fabrication routes, a possibility of their integration with other elements of an electronic circuit, highly reproducible performance, and temporal stability of the molecular junctions, including the stability against bias cycling. Such parameters as the operation voltage, which in most cases is comparable with that of the CMOS-based devices (see Table 2), and the frequency range, which is still highly unexplored for most of the molecular rectifiers, are important as well.

Another important issue is the practically accessible size of molecular rectifiers. Along with the fundamental science and novel technological approaches, the miniaturization of functional elements of electronic circuits was a starting point and is a driving force of molecular electronics. The ultimate miniaturization is provided by single-molecule devices, which, however, feature comparably low $RR$ values on the order of $10^3$ at the most as well as a limited integration capability with the CMOS-based elements and presumably high production costs. The more effective (in context of $RR$, see Table 2) SAM-, multilayer-, and CP/MOF-based rectifiers represent molecular assembles, with few nm vertical size but a non-precisely determined lateral size, experimentally defined by the lateral dimensions of the contact area by the top electrode. Apart from the related fabrication and technological issues, it is a good question how far the lateral size of such rectifiers can be reduced to still have their stable and reliable performance.

The fundamental basis for molecular rectification is a rational design of relative molecular systems relying on a deep understanding of the relationship between the molecular structure and charge transport behavior, which has not yet been completely achieved. Such an understanding will require significant joint efforts by both experiment and theory, with a variety of custom-designed test and reference systems as well as related theoretical models and computer simulations. A further related issue is the coupling of molecules to the interfaces, with a variety of bonding motifs and specific interactions, affected both by molecular orientation and morphology of the electrodes being involved. It can be in particular useful to utilize the more stable carbene bonding to gold electrodes as compared to less stable and oxidation-prone thiolate anchor,[162] which can contribute positively to the reliability, reproducibility, and stability of the molecular devices.

**Table 2. Comparison of selected molecular diodes with conventional semiconductor-based devices.**

| Diode | Type | Measurement approach | Rectification ratio (RR) | Refs |
|---|---|---|---|---|
| *Molecular-based diodes* | | | | |
| 1,8-nonodiyne | single molecule | STM break-junction | $> 4.0 \times 10^3$ | 70 |
| Ru-polypyridyl complexes | single molecule | CP-AFM | $10^{3 \pm 0.6}$ ($\pm 0.9$ V) | 73 |
| Fc-substututed alkanethiolates on Ag | SAM | two-terminal junction | $1.0 \times 10^2$ ($\pm 1$ V) | 96 |
| decanethiolate terminated with fullerene | SAM | CP-AFM | 940 ($\pm 1$ V) | 99 |
| $S(CH_2)_{11} MV_2^+ X^-_2$ | SAM | two-terminal junction | $2.5 \times 10^4$ ($\pm 1$ V) | 53 |
| alkanethiolates with the Fc□C□C□Fc termini on Pt | SAM | two-terminal junction | $6.3 \times 10^5$ ($\pm 3$ V) | 100 |
| (OPT2)/$MoS_2$ | heterostructure | CP-AFM | $1.79 \times 10^3$ ($\pm 1$V) | 144 |
| $[Cd(4-bpd)(SCN)_2]_n$ | MOF | two-terminal junction | 86.48 ($\pm 5$ V) | 155 |
| Cu (II)-BTEC CPs doped with TCNQ | CP | two-terminal junction | $>10^5$ ($\pm 5$V) | 157 |
| $Cu_2(adc)_2(dabco)$ and $C_{60}@Cu_2(bdc)_2(dabco)$ | SURMOFs | two-terminal junction | $\sim 10^6$ ($\pm 1$V) | 162 |
| (Fe-tpy-DTE-tpy)$_n$ | multilayer | CP-AFM | $2.0 \times 10^2$ ($\pm 1$ V) | 128 |
| $C_{60}$ and pentacene bilayer | multilayer | STM | $10^3$ ($\pm 1$V) | 124 |
| Dioctylbenzothienobenzo-thiophene | multilayer | two-terminal junction | $10^5$ ($\pm 2$V) | 135 |
| *Semiconductor-based diodes* | | | | |
| 1N4007 | | two probe measurements | $\geq 10^5$ ($\pm 1$V) | 157 |
| 1N4733A | | two probe measurements | $\geq 10^5$ ($\pm 1$V) | 157 |
| silicon-based diode | | two probe measurements | $\sim 10^7$ ($\pm 2$V) | 167,168 |

Fc = ferrocene; MV = methyl viologen; X⁻ = counter ion; 4-bpd = 1,4-bis(4- pyridyl)-2,3-diaza-1,3-butadiene; TCNQ = 7,7,8,8-tetracyanoquinodimethane; BTEC = 1,2,4,5-benzenetetracarboxylic acid; OPT2 = biphenyl-4-monothiol; adc = 9,10-anthracene dicarboxylate; bdc = 1,4-benzene dicarboxylate; dabco = 1,4-diazabicyclo[2.2.2]octane; Fe-tpy-DTE-tpy = iron(II) bis-terpyridine-appended dithienylethene.

# 9. Outlook

It has been broadly forecasted that silicon-based electronics will soon reach its scalability limit, and molecular-based electronic circuitry can become a valuable addition or even a potential alternative to the standard devices, bearing numerous advantages that outweigh the molecular-specific drawbacks. Generally, a successor for the existing CMOS technologies should not only provide the same specifications as International Technology Roadmap for Semiconductors (IRTS) and non-IRTS anticipated CMOS nodes but also should outperform the prevalent technology in at least one of the key aspects, such as power consumption, fabrication costs, and/or performance. Consequently, molecular-based circuitries must be reliable, durable, fast responding, and capable of performing complex electronic functions, apart from the production cost factor, which is always of importance for commercialization of any product. So, it is still a long way for molecular electronics to mature into a competitive technology, with multiple intermediate steps and milestones. Important objectives and steps in this context, addressed in more detail in the previous section, are a deep understanding of the relationship between the molecular structure and charge transport properties allowing a rational design of molecular rectifiers, better control of the molecule/electrode interfaces, developing technologies for large-scale manufacturing of reliable devices, and ensuring their robust performance and stability. This will require significant experimental efforts and progress in theory by the scientific community but also noticeable involvement of the industrial community, in context of specific technologies merging the conventual CMOS and molecular building blocks as well as of the fabrication of (prototype) devices relying on molecular electronics only. Because of the complexity of the problem and the involvement of many factors and players, it is difficult to say at the moment what particular approach to molecular rectification will first reach the capability for practical, commercial application. The progress achieved so far let us, however, hope that, along with organic semiconductor rectifiers[169–171], molecular rectifiers will become a useful and important part of future micro to nanoelectronics technology.

**Acknowledgements**: PCM acknowledges the Department of Science and Technology for start-up research grant (SRG/2019/000391), IIT Kanpur (IITK/CHM/2019044) for initiation and special grant, and the Council of Scientific & Industrial Research (CSIR, Sanctioned NO.:01(3049)/21/EMR-II), New Delhi.

**Author contributions**: J.A.F., M.Z. and P.C.M. conceived the manuscript; M.Z. and P.C.M. made its final editing, preparing it for the submission. All authors contributed to the writing of the manuscript and its subsequent revisions and editing. All authors agreed for the submission of the manuscript.

**Competing interests**: The authors declare no competing interests.

# References

1. Herrer, L., Martín, S. & Cea, P. Nanofabrication techniques in large-area molecular electronic devices. *Appl. Sci.* **10**, (2020).

2. Xiang, D., Wang, X., Jia, C., Lee, T. & Guo, X. Molecular-Scale Electronics: From Concept to Function. *Chem. Rev.* **116**, 4318–4440 (2016).

3. Bergren, A. J., Zeer-wanklyn, L. & Semple, M. Musical molecules : the molecular junction as an active component in audio distortion circuits. *J. Phys. Condens. Matter* **28**, 094011 (2016).

4. Vilan, A. & Cahen, D. Chemical Modification of Semiconductor Surfaces for Molecular Electronics. *Chem. Rev.* **117**, 4624–4666 (2017).

5. Kong, G. D. *et al.* Mixed Molecular Electronics: Tunneling Behaviors and Applications of Mixed Self-Assembled Monolayers. *Adv. Electron. Mater.* **6**, 1901157 (2020).

6. Liu, Y., Qiu, X., Soni, S. & Chiechi, R. C. Charge transport through molecular ensembles: Recent progress in molecular electronics. *Chem. Phys. Rev.* **2**, 021303 (2021).

7. Fu, T., Zang, Y., Zou, Q., Nuckolls, C. & Venkataraman, L. Using deep learning to identify molecular junction characteristics. *Nano Lett.* **20**, 3320–3325 (2020).

8. Liao, K. C., Hsu, L. Y., Bowers, C. M., Rabitz, H. & Whitesides, G. M. Molecular Series-Tunneling Junctions. *J. Am. Chem. Soc.* **137**, 5948–5954 (2015).

9. Belding, L. *et al.* Conformation, and Charge Tunneling through Molecules in SAMs. *J. Am. Chem. Soc.* **143**, 3481–3493 (2021).

10. Liu, L. *et al.* Enhancement of Intrinsic Proton Conductivity and Aniline Sensitivity by Introducing Dye Molecules into the MOF Channel. *ACS Appl. Mater. Interfaces* **11**, 16490–16495 (2019).

11. Vilan, A., Aswal, D. & Cahen, D. Large-Area, Ensemble Molecular Electronics: Motivation and Challenges. *Chem. Rev.* **117**, 4248–4286 (2017).

12. Kumar, S. *et al.* Chemical Locking in Molecular Tunneling Junctions Enables Nonvolatile Memory with Large On–Off Ratios. *Adv. Mater.* **31**, 1807831 (2019).

13. Goswami, S. *et al.* Robust resistive memory devices using solution-processable metal-coordinated azo aromatics. *Nat. Mater.* **16**, 1216–1224 (2017).

14. Li, Y. *et al.* Gate controlling of quantum interference and direct observation of anti-resonances in single molecule charge transport. *Nat. Mater.* **18**, 357–363 (2019).

15. Diez-Perez, I. *et al.* Controlling single-molecule conductance through lateral coupling of π orbitals. *Nat. Nanotechnol.* **6**, 226–231 (2011).

16. Ho Choi, S., Kim, B. & Frisbie, C. D. Electrical Resistance of Long Conjugated Molecular Wires. *Science (80-. ).* **320**, 1482–1486 (2008).

17. Gupta, R. *et al.* Electrochemical Potential-Driven High-Throughput Molecular Electronic and Spintronic Devices: From Molecules to Applications. *Angew. Chemie Int. Ed.* **60**, 26904–26921 (2021).

18. Bonifas, A. P. & McCreery, R. L. Soft Au, Pt and Cu contacts for molecular junctions through surface-diffusion-mediated deposition. *Nat. Nanotechnol.* **5**, 612–617 (2010).

19. Love, J. C., Estroff, L. A., Kriebel, J. K., Nuzzo, R. G. & Whitesides, G. M. Self-Assembled Monolayers of Thiolates on Metals as a Form of Nanotechnology. *Chem. Rev.* **105**, 1103–1170 (2005).

20. Venkataraman, L., Klare, J. E., Nuckolls, C., Hybertsen, M. S. & Steigerwald, M. L. Dependence of

single-molecule junction conductance on molecular conformation. *Nature* **442**, 904–907 (2006).

21. Jash, P., Parashar, R. K., Fontanesi, C. & Mondal, P. C. The Importance of Electrical Impedance Spectroscopy and Equivalent Circuit Analysis on Nanoscale Molecular Electronic Devices. *Adv. Funct. Mater.* 2109956 (2021). doi:10.1002/adfm.202109956

22. Chen, X. & Nijhuis, C. A. The Unusual Dielectric Response of Large Area Molecular Tunnel Junctions Probed with Impedance Spectroscopy. *Adv. Electron. Mater.* 2100495 (2021).

23. Aviram, A. & Ratner, M. A. Molecular rectifiers. *Chem. Phys. Lett.* **29**, 277–283 (1974).

24. Perrin, M. L. *et al.* Single-Molecule Resonant Tunneling Diode. *J. Phys. Chem. C* **119**, 5697–5702 (2015).

25. Ng, M. K., Lee, D. C. & Yu, L. Molecular diodes based on conjugated diblock co-oligomers. *J. Am. Chem. Soc.* **124**, 11862–11863 (2002).

26. Díez-Pérez, I. *et al.* Rectification and stability of a single molecular diode with controlled orientation. *Nat. Chem.* **1**, 635–641 (2009).

27. Li, Y., Yao, J., Liu, C. & Yang, C. Theoretical investigation on electron transport properties of a single molecular diode. *J. Mol. Struct. THEOCHEM* **867**, 59–63 (2008).

28. Handayani, M., Gohda, S., Tanaka, D. & Ogawa, T. Design and synthesis of perpendicularly connected metal porphyrin-imide dyads for two-terminal wired single molecular diodes. *Chem. - A Eur. J.* **20**, 7655–7664 (2014).

29. Liu, R., Ke, S. H., Yang, W. & Baranger, H. U. Organometallic molecular rectification. *J. Chem. Phys.* **124**, (2006).

30. Troisi, A. & Ratner, M. A. Conformational molecular rectifiers. *Nano Lett.* **4**, 591–595 (2004).

31. García, M., Guadarrama, P., Ramos, E. & Fomine, S. Rectifying behavior of [60]fullerene charge transfer complexes: A theoretical study. *Synth. Met.* **161**, 2390–2396 (2011).

32. Elbing, M. *et al.* A single-molecule diode. *Proc. Natl. Acad. Sci. U. S. A.* **102**, 8815–8820 (2005).

33. Xie, Z., Bâldea, I. & Frisbie, C. D. Why one can expect large rectification in molecular junctions based on alkane monothiols and why rectification is so modest. *Chem. Sci.* **9**, 4456–4467 (2018).

34. Hnid, I. *et al.* Unprecedented ON/OFF Ratios in Photoactive Diarylethene-Bisthienylbenzene Molecular Junctions. *Nano Lett.* **21**, 7555–7560 (2021).

35. Nijhuis, C. A., Reus, W. F., Siegel, A. C. & Whitesides, G. M. A Molecular Half-Wave Rectifier. *J. Am. Chem. Soc.* **133**, 15397–15411 (2011).

36. Hipps, K. W. MOLECULAR ELECTRONICS: It's All About Contacts. *Science (80-. ).* **294**, 536–537 (2001).

37. Reichert, J. *et al.* Driving current through single organic molecules. *Phys. Rev. Lett.* **88**, 176804 (2002).

38. Shpaisman, H. *et al.* Electronic contact deposition onto organic molecular monolayers: Can we detect metal penetration? *Adv. Funct. Mater.* **20**, 2181–2188 (2010).

39. Kushmerick, J. G. *et al.* Metal-Molecule Contacts and Charge Transport across Monomolecular Layers: Measurement and Theory. *Phys. Rev. Lett.* **89**, 086802 (2002).

40. Taylor, J., Brandbyge, M. & Stokbro, K. Theory of Rectification in Tour Wires: The Role of Electrode Coupling. *Phys. Rev. Lett.* **89**, 138301 (2002).

41. Metzger, R. M. Unimolecular Electronics. *Chem. Rev.* **115**, 5056–5115 (2015).

42. Batra, A. *et al.* Tuning Rectification in Single-Molecular Diodes. *Nano Lett.* **22**, 6233–6237 (2013).

43. Ding, W., Negre, C. F. A., Vogt, L. & Batista, V. S. Single Molecule Rectification Induced by the Asymmetry of a Single Frontier Orbital. *J. Chem. Theory Comput.* **10**, 3393–3400 (2014).

44. Van Dyck, C. & Ratner, M. A. Molecular rectifiers: A new design based on asymmetric anchoring moieties. *Nano Lett.* **15**, 1577–1584 (2015).

45. Zhang, G., Ratner, M. A. & Reuter, M. G. Is molecular rectification caused by asymmetric electrode couplings or by a molecular bias drop? *J. Phys. Chem. C* **119**, 6254–6260 (2015).

46. Gates, B. D. *et al.* New approaches to nanofabrication: Molding, printing, and other techniques. *Chem. Rev.* **105**, 1171–1196 (2005).

47. Whitesides, G. M., Kriebel, J. K. & Love, J. C. Molecular engineering of Surfaces Using Self-Assembled Monolayers. *Sci. Prog.* **88**, 17–48 (2005).

48. Bruinink, C. M. *et al.* Supramolecular microcontact printing and dip-pen nanolithography on molecular printboards. *Chem. - A Eur. J.* **11**, 3988–3996 (2005).

49. Kim, T., Liu, Z.-F., Lee, C., Neaton, J. B. & Venkataraman, L. Charge transport and rectification in molecular junctions formed with carbon-based electrodes. *Proc. Natl. Acad. Sci.* **111**, 10928–10932 (2014).

50. Stokbro, K., Taylor, J. & Brandbyge, M. Do Aviram−Ratner Diodes Rectify? *J. Am. Chem. Soc.* **125**, 3674–3675 (2003).

51. Yuan, L., Breuer, R., Jiang, L., Schmittel, M. & Nijhuis, C. A. A Molecular Diode with a Statistically Robust Rectification Ratio of Three Orders of Magnitude. *Nano Lett.* **15**, 5506–5512 (2015).

52. Chen, X. *et al.* Molecular diodes with rectification ratios exceeding 10 5 driven by electrostatic interactions. *Nat. Nanotech.* **12**, 797–803 (2017).

53. Han, Y. *et al.* Electric-field-driven dual-functional molecular switches in tunnel junctions. *Nat. Mater.* **19**, 843–848 (2020).

54. Thompson, D., Barco, E. Del & Nijhuis, C. A. Design principles of dual-functional molecular switches in solid-state tunnel junctions. *Appl. Phys. Lett.* **117**, 030502 (2020).

55. Wang, L., Wang, L., Zhang, L. & Xiang, D. Advance of Mechanically Controllable Break Junction for Molecular Electronics. *Top. Curr. Chem.* **375**, 61 (2017).

56. Gehring, P., Thijssen, J. M. & van der Zant, H. S. J. Single-molecule quantum-transport phenomena in break junctions. *Nat. Rev. Phys.* **1**, 381–396 (2019).

57. Caneva, S. *et al.* Mechanically controlled quantum interference in graphene break junctions. *Nat. Nanotechnol.* **13**, 1126–1131 (2018).

58. Bellec, A., Lagoute, J. & Repain, V. Molecular electronics: Scanning tunneling microscopy and single-molecule devices. *Comptes Rendus Chim.* **21**, 1287–1299 (2018).

59. Bouvron, S. *et al.* Charge transport in a single molecule transistor probed by scanning tunneling microscopy. *Nanoscale* **10**, 1487–1493 (2018).

60. Wold, D. J. & Frisbie, C. D. Formation of Metal - Molecule - Metal Tunnel Junctions : Microcontacts to Alkanethiol Monolayers with a Conducting AFM Tip. *J. Am. Chem. Soc.* **122**, 2970–2971 (2000).

61. Wold, D. J., Frisbie, C. D. & January, R. V. Fabrication and Characterization of Metal - Molecule - Metal Junctions by Conducting Probe Atomic Force Microscopy. *J. Am. Chem. Soc.* **123**, 5549–5556 (2001).

62. Kim, Beebe, J. M., Jun, Y., Zhu, X.-Y. & Frisbie, C. D. Correlation between HOMO Alignment and

Contact Resistance in Molecular Junctions: Aromatic Thiols versus Aromatic Isocyanides. *J. Am. Chem. Soc.* **128**, 4970–4971 (2006).

63. Pasupathy, A. N. *et al.* The Kondo effect in the presence of ferromagnetism. *Science (80-. ).* **306**, 86–89 (2004).

64. Vazquez, H. *et al.* Probing the conductance superposition law in single-molecule circuits with parallel paths. *Nat. Nanotechnol.* **7**, 663–667 (2012).

65. Pramod Reddy, Sung-Yeon Jang, Rachel A. Segalman, A. M. Thermoelectricity in Molecular Juncrions. *Science (80-. ).* **315**, 1568–1571 (2007).

66. Rodriguez-Gonzalez, S. *et al.* HOMO Level Pinning in Molecular Junctions: Joint Theoretical and Experimental Evidence. *J. Phys. Chem. Lett.* **9**, 2394–2403 (2018).

67. Xin, N. *et al.* Concepts in the design and engineering of single-molecule electronic devices. *Nat. Rev. Phys.* **1**, 211–230 (2019).

68. Capozzi, B. *et al.* Single-molecule diodes with high rectification ratios through environmental control. *Nat. Nanotechnol.* **10**, 522–527 (2015).

69. Iwane, M., Fujii, S. & Kiguchi, M. Molecular Diode Studies Based on a Highly Sensitive Molecular Measurement Technique. *Sensors* **17**, 956 (2017).

70. Aragonès, A. C. *et al.* Single-molecule electrical contacts on silicon electrodes under ambient conditions. *Nat. Commun.* **8**, 15056 (2017).

71. Grozema, F. & Zant, H. S. J. Van Der. A gate-tunable single-molecule diode. *Nanoscale* **8**, 8919–8923 (2016).

72. Schwarz, F. *et al.* Field-induced conductance switching by charge-state alternation in organometallic single-molecule junctions. *Nat. Nanotechnol.* **11**, 170–176 (2016).

73. Atesci, H. *et al.* Humidity-controlled rectification switching in ruthenium-complex molecular junctions. *Nat. Nanotechnol.* **13**, 117–121 (2018).

74. Xu, K. & Xie, S. Self-assembled molecular devices: a minireview. *Instrum. Sci. Technol.* **48**, 86–111 (2020).

75. Schmaltz, T., Sforazzini, G., Reichert, T. & Frauenrath, H. Self-Assembled Monolayers as Patterning Tool for Organic Electronic Devices. *Adv. Mater.* **29**, 1605286 (2017).

76. Singh, M., Kaur, N. & Comini, E. The role of self-assembled monolayers in electronic devices. *J. Mater. Chem. C* **8**, 3938–3955 (2020).

77. Casalini, S., Bortolotti, C. A., Leonardi, F. & Biscarini, F. Self-assembled monolayers in organic electronics. *Chem. Soc. Rev.* **46**, 40–71 (2017).

78. Otero, R., Gallego, J. M., De Parga, A. L. V., Martín, N. & Miranda, R. Molecular self-assembly at solid surfaces. *Adv. Mater.* **23**, 5148–5176 (2011).

79. Ulman, A. Formation and Structure of Self-Assembled Monolayers. *Chem. Rev.* **96**, 1533–1554 (1996).

80. Vericat, C. *et al.* Surface characterization of sulfur and alkanethiol self-assembled monolayers on Au(111). *J. Phys. Condens. Matter* **18**, (2006).

81. Pujari, S. P., Scheres, L., Marcelis, A. T. M. & Zuilhof, H. Covalent surface modification of oxide surfaces. *Angew. Chemie - Int. Ed.* **53**, 6322–6356 (2014).

82. Paniagua, S. A. *et al.* Phosphonic Acids for Interfacial Engineering of Transparent Conductive Oxides. *Chem. Rev.* **116**, 7117–7158 (2016).

83. Mondal, P. C., Singh, V. & Zharnikov, M. Nanometric Assembly of Functional Terpyridyl Complexes on Transparent and Conductive Oxide Substrates: Structure, Properties, and Applications. *Acc. Chem. Res.* **50**, 2128–2138 (2017).

84. Fabre, B. Functionalization of Oxide-Free Silicon Surfaces with Redox-Active Assemblies. *Chem. Rev.* **116**, 4808–4849 (2016).

85. Peng, W. *et al.* Silicon Surface Modification and Characterization for Emergent Photovoltaic Applications Based on Energy Transfer. *Chem. Rev.* **115**, 12764–12796 (2015).

86. Singh, V., Mondal, P. C., Singh, A. K. & Zharnikov, M. Molecular sensors confined on $SiO_x$ substrates. *Coord. Chem. Rev.* **330**, 144–163 (2017).

87. Yuan, L. *et al.* Controlling the direction of rectification in a molecular diode. *Nat. Commun.* **6**, 6324 (2015).

88. Yuan, L., Thompson, D., Cao, L., Nerngchangnong, N. & Nijhuis, C. A. One Carbon Matters: The Origin and Reversal of Odd–Even Effects in Molecular Diodes with Self-Assembled Monolayers of Ferrocenyl-Alkanethiolates. *J. Phys. Chem. C* **119**, 17910–17919 (2015).

89. Wang, L. *et al.* Unraveling the Failure Modes of Molecular Diodes: The Importance of the Monolayer Formation Protocol and Anchoring Group to Minimize Leakage Currents. *J. Phys. Chem. C* **123**, 19759–19767 (2019).

90. Starr, R. L. *et al.* Gold–Carbon Contacts from Oxidative Addition of Aryl Iodides. *J. Am. Chem. Soc.* **142**, 7128–7133 (2020).

91. Haj-Yahia, A. E. *et al.* Substituent variation drives metal/monolayer/semiconductor junctions from strongly rectifying to ohmic behavior. *Adv. Mater.* **25**, 702–706 (2013).

92. Nguyen, Q. Van, Xie, Z. & Daniel Frisbie, C. Quantifying Molecular Structure-Tunneling Conductance Relationships: Oligophenylene Dimethanethiol vs Oligophenylene Dithiol Molecular Junctions. *J. Phys. Chem. C* **125**, 4292–4298 (2021).

93. Vilan, A., Ghabboun, J. & Cahen, D. Molecule-metal polarization at rectifying GaAs interfaces. *J. Phys. Chem. B* **107**, 6360–6376 (2003).

94. Sachan, P. & Mondal, P. C. Versatile electrochemical approaches towards the fabrication of molecular electronic devices. *Analyst* **145**, 1563–1582 (2020).

95. Cademartiri, L. *et al.* Electrical Resistance of $Ag^{TS}–S(CH_2)_{n−1}CH_3//Ga_2O_3/EGaIn$ Tunneling Junctions. *J. Phys. Chem. C* **116**, 10848–10860 (2012).

96. Nijhuis, C. A., Reus, W. F. & Whitesides, G. M. Molecular rectification in metal-SAM-metal oxide-metal junctions. *J. Am. Chem. Soc.* **131**, 17814–17827 (2009).

97. Nijhuis, C. A., Reus, W. F. & Whitesides, G. M. Mechanism of Rectification in Tunneling Junctions Based on Molecules with Asymmetric Potential Drops. *J. Am. Chem. Soc.* **132**, 18386–18401 (2010).

98. Nijhuis, C. A., Reus, W. F., Barber, J. R., Dickey, M. D. & Whitesides, G. M. Charge transport and rectification in arrays of SAM-based tunneling junctions. *Nano Lett.* **10**, 3611–3619 (2010).

99. Qiu, L. *et al.* Rectification of current responds to incorporation of fullerenes into mixed-monolayers of alkanethiolates in tunneling junctions. *Chem. Sci.* **8**, 2365–2372 (2017).

100. Chen, X. *et al.* Molecular diodes with rectification ratios exceeding 105 driven by electrostatic interactions. *Nat. Nanotechnol.* **12**, 797–803 (2017).

101. Yuan, L. *et al.* Transition from direct to inverted charge transport Marcus regions in molecular junctions via molecular orbital gating. *Nat. Nanotechnol.* **13**, 322–329 (2018).

102. Han, Y. *et al.* Reversal of the Direction of Rectification Induced by Fermi Level Pinning at Molecule-Electrode Interfaces in Redox-Active Tunneling Junctions. *ACS Appl. Mater. Interfaces* **12**, 55044–55055 (2020).

103. Park, J. *et al.* Rectification in Molecular Tunneling Junctions Based on Alkanethiolates with Bipyridine-Metal Complexes. *J. Am. Chem. Soc.* **143**, 2156–2163 (2021).

104. Kong, G. D. *et al.* Interstitially Mixed Self-Assembled Monolayers Enhance Electrical Stability of Molecular Junctions. *Nano Lett.* **21**, 3162–3169 (2021).

105. Fereiro, J. A., Bendikov, T., Pecht, I., Sheves, M. & Cahen, D. Protein binding and orientation matter: Bias-induced conductance switching in a mutated azurin junction. *J. Am. Chem. Soc.* **142**, 19217–19225 (2020).

106. Rinaldi, R. *et al.* Solid-state molecular rectifier based on self-organized metalloproteins. *Adv. Mater.* **14**, 1453–1457 (2002).

107. Lewis, F. D. *et al.* Direct measurement of hole transport dynamics in DNA. *Nature* **406**, 51–53 (2000).

108. Giese, B., Amaudrut, J., Köhler, A. K., Spormann, M. & Wessely, S. Direct observation of hole transfer through DNA by hopping between adenine bases and by tunnelling. *Nature* **412**, 318–320 (2001).

109. Genereux, J. C. & Barton, J. K. DNA charges ahead. *Nat. Chem.* **1**, 106–107 (2009).

110. Mallajosyula, S. S. & Pati, S. K. Toward DNA conductivity: A theoretical perspective. *J. Phys. Chem. Lett.* **1**, 1881–1894 (2010).

111. Livshits, G. I. *et al.* Long-range charge transport in single G-quadruplex DNA molecules. *Nat. Nanotechnol.* **9**, 1040–1046 (2014).

112. Arya, S. K., Solanki, P. R., Datta, M. & Malhotra, B. D. Recent advances in self-assembled monolayers based biomolecular electronic devices. *Biosens. Bioelectron.* **24**, 2810–2817 (2009).

113. Panda, S. S., Katz, H. E. & Tovar, J. D. Solid-state electrical applications of protein and peptide based nanomaterials. *Chem. Soc. Rev.* **47**, 3640–3658 (2018).

114. Wang, K. *et al.* Structure determined charge transport in single DNA molecule break junctions. *Chem. Sci.* **5**, 3425–3431 (2014).

115. Liu, S. *et al.* Direct conductance measurement of individual metallo-DNA duplexes within single-molecule break junctions. *Angew. Chemie - Int. Ed.* **50**, 8886–8890 (2011).

116. Hihath, J., Guo, S., Zhang, P. & Tao, N. Effects of cytosine methylation on DNA charge transport. *J. Phys. Condens. Matter* **24**, (2012).

117. Guo, C. *et al.* Molecular rectifier composed of DNA with high rectification ratio enabled by intercalation. *Nat. Chem.* **8**, 484–490 (2016).

118. Oliveira, J. I. N., Albuquerque, E. L., Fulco, U. L., Mauriz, P. W. & Sarmento, R. G. Electronic transport through oligopeptide chains: An artificial prototype of a molecular diode. *Chem. Phys. Lett.* **612**, 14–19 (2014).

119. Feng, M. *et al.* Recent Advances in Multilayer-Structure Dielectrics for Energy Storage Application. *Adv. Sci.* **8**, 2102221 (2021).

120. Hwang, S., Lee, E. J., Song, D. & Jeong, N. C. High Proton Mobility with High Directionality in Isolated Channels of MOF-74. *ACS Appl. Mater. Interfaces* **10**, 35354–35660 (2018).

121. Li, C. *et al.* Recent development and applications of electrical conductive MOFs. *Nanoscale* **13**, 485–

509 (2021).

122. Wang, M., Dong, R. & Feng, X. Two-dimensional conjugated metal-organic frameworks (2Dc-MOFs): chemistry and function for MOFtronics. *Chem. Soc. Rev.* **50**, 2764–2793 (2021).

123. Skorupskii, G. & Dincă, M. Electrical Conductivity in a Porous, Cubic Rare-Earth Catecholate. *J. Am. Chem. Soc.* **142**, 6920–6924 (2020).

124. Smerdon, J. A., Giebink, N. C., Guisinger, N. P., Darancet, P. & Guest, J. R. Large Spatially Resolved Rectification in a Donor–Acceptor Molecular Heterojunction. *Nano Lett.* **16**, 2603–2607 (2016).

125. Tyagi, P. Multilayer edge molecular electronics devices: a review. *J. Mater. Chem.* **21**, 4733 (2011).

126. Sergi Lopes, C., Merces, L., de Oliveira, R. F., de Camargo, D. H. S. & Bof Bufon, C. C. Rectification ratio and direction controlled by temperature in copper phthalocyanine ensemble molecular diodes. *Nanoscale* **12**, 10001–10009 (2020).

127. Li, T. *et al.* Integrated molecular diode as 10 MHz half-wave rectifier based on an organic nanostructure heterojunction. *Nat. Commun.* **11**, 1–10 (2020).

128. Hnid, I. *et al.* Combining Photomodulation and Rectification in Coordination Molecular Wires Based on Dithienylethene Molecular Junctions. *J. Phys. Chem. C* **124**, 26304–26309 (2020).

129. Chandra Mondal, P., Tefashe, U. M. & McCreery, R. L. Internal Electric Field Modulation in Molecular Electronic Devices by Atmosphere and Mobile Ions. *J. Am. Chem. Soc.* **140**, 7239–7247 (2018).

130. Smith, S. R. & McCreery, R. L. Photocurrent, Photovoltage, and Rectification in Large-Area Bilayer Molecular Electronic Junctions. *Adv. Electron. Mater.* **4**, 1800093 (2018).

131. James, D. D., Bayat, A., Smith, S. R., Lacroix, J. C. & McCreery, R. L. Nanometric building blocks for robust multifunctional molecular junctions. *Nanoscale Horizons* **3**, 45–52 (2018).

132. Seo, S. *et al.* Recent Progress in Artificial Synapses Based on Two-Dimensional van der Waals Materials for Brain-Inspired Computing. *ACS Appl. Electron. Mater.* **2**, 371–388 (2020).

133. Wu, C., Kim, T. W., Choi, H. Y., Strukov, D. B. & Yang, J. J. Flexible three-dimensional artificial synapse networks with correlated learning and trainable memory capability. *Nat. Commun.* **8**, 752 (2017).

134. Gupta, R., Jash, P. & Mondal, P. C. Nanoscale molecular layers for memory devices: challenges and opportunities for commercialization. *J. Mater. Chem. C* **9**, 11497–11516 (2021).

135. Yang, C. *et al.* An Optically Modulated Organic Schottky-Barrier Planar-Diode-Based Artificial Synapse. *Adv. Opt. Mater.* **8**, 2000153 (2020).

136. Xie, Z., Bâldea, I., Smith, C. E., Wu, Y. & Frisbie, C. D. Experimental and Theoretical Analysis of Nanotransport in Oligophenylene Dithiol Junctions as a Function of Molecular Length and Contact Work Function. *ACS Nano* **9**, 8022–8036 (2015).

137. Kilgour, M. & Segal, D. Charge transport in molecular junctions: From tunneling to hopping with the probe technique. *J. Chem. Phys.* **143**, (2015).

138. Thomas, J. O. *et al.* Understanding resonant charge transport through weakly coupled single-molecule junctions. *Nat. Commun.* **10**, 1–9 (2019).

139. Liu, J. *et al.* A rectifying diode with hysteresis effect from an electroactive hybrid of carbazole-functionalized polystyrene with CdTe nanocrystals via electrostatic interaction. *Sci. China Chem.* **53**, 2324–2328 (2010).

140. Wang, L. *et al.* Rectification-Regulated Memristive Characteristics in Electron-Type CuPc-Based Element for Electrical Synapse. *Adv. Electron. Mater.* **3**, 1–8 (2017).

141. Lu, H. *et al.* Polymer-carbon dot hybrid structure for a self-rectifying memory device by energy level offset and doping. *RSC Adv.* **8**, 13917–13920 (2018).

142. Mohanta, K., Batabyal, S. K. & Pal, A. J. Organization of organic molecules with inorganic nanoparticles: Hybrid nanodiodes. *Adv. Funct. Mater.* **18**, 687–693 (2008).

143. Banerjee, A., Kundu, B. & Pal, A. J. Hybrid heterojunctions between a 2D transition metal dichalcogenide and metal phthalocyanines: Their energy levels: Vis-à-vis current rectification. *Phys. Chem. Chem. Phys.* **19**, 28450–28457 (2017).

144. Shin, J. *et al.* Tunable rectification in a molecular heterojunction with two-dimensional semiconductors. *Nat. Commun.* **11**, 1–7 (2020).

145. Garg, S. *et al.* Conductance Photoswitching of Metal–Organic Frameworks with Embedded Spiropyran. *Angew. Chemie Int. Ed.* **58**, 1193–1197 (2019).

146. Sheberla, D. *et al.* Conductive MOF electrodes for stable supercapacitors with high areal capacitance. *Nat. Mater.* **16**, 220–224 (2017).

147. Dou, J.-H. *et al.* Atomically precise single-crystal structures of electrically conducting 2D metal–organic frameworks. *Nat. Mater.* **20**, 222–228 (2021).

148. Wang, M., Dong, R. & Feng, X. Two-dimensional conjugated metal-organic frameworks (2Dc-MOFs): chemistry and function for MOFtronics. *Chem. Soc. Rev.* **50**, 2764–2793 (2021).

149. Johnson, E. M., Ilic, S. & Morris, A. J. Design Strategies for Enhanced Conductivity in Metal-Organic Frameworks. *ACS Central Science* **7**, 445–453 (2021).

150. Neumann, T. *et al.* Superexchange Charge Transport in Loaded Metal Organic Frameworks. *ACS Nano* **10**, 7085–7093 (2016).

151. Liu, J. *et al.* Electric Transport Properties of Surface-Anchored Metal–Organic Frameworks and the Effect of Ferrocene Loading. *ACS Appl. Mater. Interfaces* **7**, 9824–9830 (2015).

152. Meng, X., Wang, H. N., Wang, L. S., Zou, Y. H. & Zhou, Z. Y. Enhanced proton conductivity of a MOF-808 framework through anchoring organic acids to the zirconium clusters by post-synthetic modification. *CrystEngComm* **21**, 3146–3150 (2019).

153. Yang, S. *et al.* Enhancing MOF performance through the introduction of polymer guests. *Coord. Chem. Rev.* **427**, 213525 (2021).

154. Xie, L. S., Skorupskii, G. & Dincă, M. Electrically Conductive Metal-Organic Frameworks. *Chem. Rev.* **120**, 8536–8580 (2020).

155. Halder, S. *et al.* A Cd(II)-based MOF as a photosensitive Schottky diode: Experimental and theoretical studies. *Dalt. Trans.* **46**, 11239–11249 (2017).

156. Ghorai, P. *et al.* The development of a promising photosensitive Schottky barrier diode using a novel Cd(II) based coordination polymer. *Dalt. Trans.* **46**, 13531–13543 (2017).

157. Prasoon, A. *et al.* Achieving current rectification ratios ≥ 105 across thin films of coordination polymer. *Chem. Sci.* **10**, 10040–10047 (2019).

158. Das, K. S. *et al.* Utilization of counter anions for charge transportation in the electrical device fabrication of Zn(ii) metal-organic frameworks. *Dalt. Trans.* **49**, 17005–17016 (2020).

159. Cao, L.-A. *et al.* A highly oriented conductive MOF thin film-based Schottky diode for self-powered light and gas detection. *J. Mater. Chem. A* **8**, 9085–9090 (2020).

160. Halder, S. *et al.* A Cd(ii)-based MOF as a photosensitive Schottky diode: experimental and theoretical studies. *Dalt. Trans.* **46**, 11239–11249 (2017).

161. Das, K. S. *et al.* Utilization of counter anions for charge transportation in the electrical device fabrication of Zn(ii) metal–organic frameworks. *Dalt. Trans.* **49**, 17005–17016 (2020).

162. Chandresh, A., Liu, X., Wöll, C. & Heinke, L. Programmed Molecular Assembly of Abrupt Crystalline Organic/Organic Heterointerfaces Yielding Metal-Organic Framework Diodes with Large On-Off Ratios. *Adv. Sci.* **8**, 2001884 (2021).

163. Goddard, W. A. *Handbook of Nanoscience, Engineering, and Technology*. (CRC Press, 2002). doi:10.1201/9781420040623

164. Mujica, V., Ratner, M. A. & Nitzan, A. Molecular rectification: Why is it so rare? *Chem. Phys.* **281**, 147–150 (2002).

165. Ma, J., Yang, C. L., Wang, L. Z., Wang, M. S. & Ma, X. G. Controllable low-bias negative differential resistance, switching, and rectifying behaviors of dipyrimidinyl-diphenyl induced by contact mode. *Phys. B Condens. Matter* **434**, 32–37 (2014).

166. Lo, E. Transport Properties of a Single-Molecule Diode. 4931–4939 (2012).

167. Szydlo, N., Chartier, E., Proust, N., Magariño, J. & Kaplan, D. High current post-hydrogenated chemical vapor deposited amorphous silicon p - i - n diodes. *Appl. Phys. Lett.* **40**, 988–990 (1982).

168. Wang, Q. *et al.* High-current-density thin-film silicon diodes grown at low temperature. *Appl. Phys. Lett.* **85**, 2122–2124 (2004).

169. Myny, K., Steudel, S., Vicca, P., Genoe, J. & Heremans, P. An integrated double half-wave organic Schottky diode rectifier on foil operating at 13.56 MHz. *Appl. Phys. Lett.* **93**, 093305 (2008).

170. Viola, F. A. *et al.* A 13.56 MHz Rectifier Based on Fully Inkjet Printed Organic Diodes. *Adv. Mater.* **32**, 2002329 (2020).

171. Trasobares, J., Vuillaume, D., Théron, D. & Clément, N. A 17 GHz molecular rectifier. *Nat. Commun.* **7**, 12850 (2016).

# Glossary:

**AR model**: AR model is D-σ-A type model proposed by Aviram and Ratner, in which two conjugated moieties, a donor and an acceptor, are connected via an insulating (non-conjugated) moiety (σ bridge).

**Atomic Force Microscopy (AFM):** AFM is a powerful technique to study the morphology and structure of surfaces with nanoscale precision. It relies on a force between the cantilever tip and sample surface; the cantilever is scanned over the surface and its force-driven deflection is recorded by a position-sensitive photodiode.

**Barrier height**: Barrier height is defined as the potential difference between the Fermi level of the electrical contact and the band edge or frontier molecular orbital (either HOMO or LUMO) populated by specific charge carriers (electrons or holes).

**Break Junctions**: Break junctions is a useful tool to investigate electrical properties of single molecules and small molecular assemblies by creating a nanometre gap between two metal electrodes, bridged by molecule/molecules.

**Conducting probe AFM (CP-AFM):** CP-AFM is an AFM operational mode used to map the local variation in electrical conductivity and topography of the sample at the nanoscale level. Along with the deflection of the cantilever, the current between the cantilever and the sample surface is measured during the scanning.

**Fermi level:** Fermi level is the highest energy level occupied by electrons in a metal at an absolute zero temperature; the density of states at the Fermi level determines the thermal and electrical conductivity of a particular material.

**Inelastic Electron Tunnelling Spectroscopy (IETS):** IETS is an analytical technique to investigate the vibrational modes of molecules. In IETS, an oxide layer with molecules adsorbed on it is put between two metal electrodes and subjected to a variable bias. The inelastic contributions to the tunnelling current, associated with the molecular vibrations, are recorded.

**Mechanically controlled break junction**: Mechanically controlled break junction is a tool to examine the mechanical and electrical properties of single molecules.

**Rectification ratio**: Rectification ratio (a numerical value) is a figure of merit for the effectiveness of rectification, usually defined as the ratio of the currents measured at the bias voltages of the same magnitude but reverse polarities.

**Scanning Tunnelling Microscopy (STM):** STM is a powerful technique to study the morphology and structure of surfaces with atomic and nanoscale precision. It utilizes quantum tunnelling between the conducting tip and sample surface; the tip is scanned over the surface and the tunnelling current is recorded (one of the acquisition modes).

**Tunneling junctions:** In electronics, a tunneling junction is defined as a thin insulating barrier between the conducting electrodes in which electrons pass across the barrier through the quantum tunnelling rather than "jumping" over it.

**Work function:** It is defined as the minimum energy (usually in eV) required to remove an electron from the bulk of a given material to a point in the vacuum immediately outside the material surface. The value of work function varies from material to material.